\pdfoutput=1

\documentclass[twocolumn,english]{revtex4}
\usepackage{times,amssymb,amsmath,graphicx}
\usepackage[bookmarks=false,pdffitwindow=false,pdfstartview={FitH}]{hyperref}
\usepackage[T1]{fontenc}
\usepackage{babel,multirow}
\usepackage{color}

\begin{document}

%%  ======================================================================  %%
%%  TO JEST  >> W A Z <<  O DLUGOSCI 70+8 ZNAKOW I MA SIE MIESCIC W LINII   %%
%%  ======================================================================  %%

\title{
  Quantum transport of Dirac fermions in selected graphene
  nanosystems away from the charge-neutrality point
}

\author{Adam Rycerz\footnote{Correspondence: 
  \href{mailto:rycerz@th.if.uj.edu.pl}{rycerz@th.if.uj.edu.pl}.}}
\affiliation{Institute for Theoretical Physics,
  Jagiellonian University, \L{}ojasiewicza 11, PL--30348 Krak\'{o}w, Poland}

\date{April 18, 2025}

\begin{abstract}
Peculiar electronic properties of graphene, including the universal
{\em dc} conductivity and the pseudodiffusive shot noise, are usually
attributed to a~small vicinity of the charge-neutrality point, away from
which electron's effective mass raises, and nanostructures in graphene
start to behave similarly to familiar Sharvin contacts in semiconducting
heterostructures.
Recently, it was pointed out that as long as abrupt potential steps
separate the sample area from the leads, some graphene-specific
features can be identified relatively far from the charge-neutrality
point.
These features include conductance reduction and shot noise
enhancement compared to the standard Sharvin values.
The purpose of this paper is twofold: 
First, we extend the previous analysis based on the effective Dirac
equation, and derive the formulas that allow the calculation of the 
arbitrary charge-transfer cumulant for doped graphene.
Second, the results of analytic considerations are compared 
with numerical simulations of quantum transport on the honeycomb
lattice, for selected nanosystems for which considerations starting
from the Dirac equation cannot be directly adapted. 
For a~wedge-shaped constriction with zigzag edges, the transport
characteristics can be tuned from graphene-specific 
({\em sub-Sharvin}) values to standard Sharvin values by varying 
the electrostatic potential profile in the narrowest section.
A~similar scenario is followed by the half-Corbino disk.
In contrast, a~circular quantum dot with two narrow openings
shows a~mixed behavior appears: the conductance is close to the
Sharvin value, while the Fano factor approaches the value
characterizing the symmetric chaotic cavity. 
Carving a~hole in the quantum dot to eliminate direct trajectories
between the openings reduces the conductance to 
sub-Sharvin value, but the Fano factor is unaffected. 
Our results suggest that experimental attempts to verify the
predictions for the sub-Sharvin transport regime should focus on
systems with relatively wide openings, where the scattering at the
sample edges is insignificant next to the scattering at the 
sample-lead interfaces. 
\end{abstract}

\maketitle

%%%%%%%%%%%%%%%%%%%%%%%%%%%%%%%%%%%%%%%%%%%%%%%%%%%%%%%%%%%%%%%%%%%%%%%%%%%%%%
%%%%%%%%%%%%%%%%%%%%%%%%%%%%%%%%%%%%%%%%%%%%%%%%%%%%%%%%%%%%%%%%%%%%%%%%%%%%%%

\section{Introduction}
There are few phenomena in nature for which the results of measurements of physical quantities are given directly by the fundamental constants of nature, leaving even the question of the actual number of fundamental constants open \cite{Duf02,Leb19}.
In the second half of the last century, two phenomena from this group were discovered and theoretically described: the quantum Hall effect \cite{And75,Kli80,Lau81,Tsu82} and the Josephson effect \cite{Jos74}, which are currently used as the basis for the standards of units of resistance and electric voltage in the SI system, i.e., the ohm \cite{Pay20} and the volt \cite{YTa15}.
The discovery of the two-dimensional allotrope of carbon, graphene, made at the beginning of the 21st century \cite{Nov05,Zha05} allowed for the improvement of the Ohm standard based on the quantum Hall effect \cite{Pay20}.
(Some peculiar features of the Josephson effect in graphene were
also pointed out \cite{Tit06,Sal23}.) 
Moreover, it turned out that several material characteristics of graphene, such as the conductivity \cite{Kat06,Kat20} or visible light absorption \cite{Nai08,Sku10}, are given by the fundamental constants or dimensionless numerical coefficients.
The sub-Poissonian shot noise (quantified by the Fano factor $F=1/3$) \cite{Two06,Dan08,Ryc09,Lai16} and the anomalous Lorentz number \cite{Yos15,Cro16,Ryc21a,YTT23} for charge-neutral graphene also can be regarded as examples of such characteristics. 
Although measurements of these quantities with metrological accuracy are not possible yet, the scientific community have undoubtedly gained unique opportunities to test a theoretical model, which is the effective two-dimensional Dirac-Weyl equation for monolayer graphene \cite{Sem84,DiV84}. 
Thus, in graphene, the theoretical perspective complements the device-oriented research avenue \cite{Zor15,Niy18,Sus18,MDu21,Had23}. 

The author and Witkowski have recently find, using the effective Dirac equation, that for doped graphene samples of highly-symmetric shapes (namely, the rectangle with smooth edges and the Corbino disk) the conductance is reduced, whereas the shot noise is amplified, comparing to standard Sharvin values \cite{Ryc21b,Ryc22}.
The reduction (or amplification) is maximal when abrupt potential steps separate the sample area from the leads; for instance, the conductance $G\approx{}(\pi/4)\,G_{\rm Sharvin}$ (with $G_{\rm Sharvin}=g_0k_FW/\pi$ \cite{Sha65,Wee88,Gla88,Bez07}, $g_0$ the conductance quantum, $k_F$ the Fermi wavenumber, and $W$ the sample width \cite{diskfoo}) for the rectangle or the narrow disk (i.e., the inner-to-outer radii ratio $R_{\rm_i}/R_{\rm o}\approx{}1$), $G\approx{}(4-\pi)\,G_{\rm Sharvin}$ for the wide-disk limit ($R_{\rm_i}/R_{\rm o}\ll{}1$); the Fano factor $F\approx{1/8}$ for the rectangle or the disk with $R_{\rm_i}/R_{\rm o}\approx{}1$, $F\approx{}0.1065$ for the disk with $R_{\rm_i}/R_{\rm o}\ll{}1$.
When the potential profile gets smoothen the above-listed {\em sub-Sharvin} values evolve towards $G\approx{}G_{\rm Sharvin}$ and the Fano factor approaches the ballistic value of $F\approx{}0$. 
Later, the discussion in analytical terms was extended on the nonzero magnetic field case \cite{Ryc23,Ryc24}, showing that in a~doped disk with $R_{\rm_i}/R_{\rm o}\approx{}1$ the vanishing conductance $G\rightarrow{}0$ (notice that in the disk geometry the edge states are absent and the current is blocked at sufficiently high field except from narrow resonances via Landau levels \cite{DaS09,Ryc10,Zen19,Sus20a,Kam21}) is accompanied by a~non-trivial value of $F\approx{}0.55$, evolving towards the poissonian limit of $F\rightarrow{}1$ for $R_{\rm_i}/R_{\rm o}\ll{}1$.

\begin{figure*}[t]
\includegraphics[width=\linewidth]{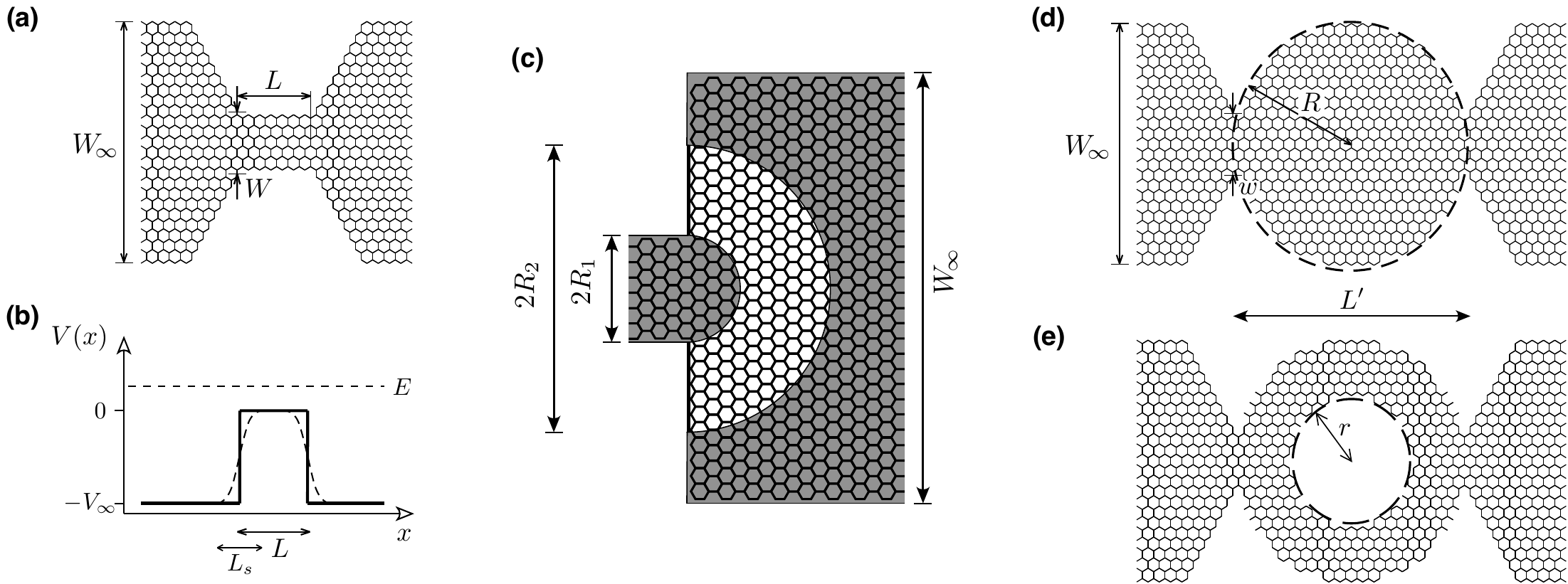}
\caption{ \label{allsetups}
  (a)--(e) Systems studied numerically in the work (schematic).
  (a) Constriction with zigzag edges containing a~narrow rectangular section
  of the width $W$ and the length $L$.
  (b) Corresponding potential profile.
  (c) Half Corbino disk (white area) with the inner radii $R_1$ and the outer
  radii $R_2$ attached to doped graphene leads with armchair edges
  (shaded areas).
  (d) Circular quantum dot of the radii $R$. The electrostatic potential
  profile (not shown) is same as in (b), but the steps are placed in the
  two narrowest sections of $w$ width at a~distance $L'$.
  (e) Circular quantum dot with a~circular hole, of the radii $r$,
  in the center and the remaining parametres same as in (d). 
}
\end{figure*}

It is the purpose of this paper to extend the discussion of sub-Sharvin transport regime in graphene by going beyond the effective Dirac equation. In particular, we address the question how reallistic (irregular) edges of a~nanosystem carved out of the honeycomb lattice affect transport characteritics? For this purpose, we perform computer simulations of quantum transport for selected systems, containing up to $336,000$ lattice sites and depicted schematically in Fig.\ \ref{allsetups}, modeled within the tight-binding Hamiltonian. The results show that sub-Sharvin characteristics are closely reconstracted for relatively short and wide systems; for longer and more complex system with multiple constrictions some less obvious scenarios (including the sub-Sharvin conductance accompanied by the shot-noise power resembling a~chaotic cavity) can be observed. 

The remaining parts of the paper are organized as follows.
In Sec.\ \ref{lanbutgra}, we present the Landauer-B\"{u}ttiker formalism for a~generic nanoscopic system and key literature results following from the effective Dirac equation for graphene at the charge-neutrality point as well as in the sub-Sharvin regime.
The analytical technique, allowing the calculation of 
higher charge-transfer cumulants for graphene at and away from the charge-neutrality point is also presented in Sec.\ \ref{lanbutgra}. 
Statistical distributions of transmission probabilities for different quantum-transport regimes, including the sub-Sharvin transport regime in graphene, are described in Sec.\ \ref{distraei}. 
The tight-binding model of graphene and our main results concerning the conductance and the Fano factor for selected nanosystems (see Fig.\ \ref{allsetups}) are presented in Sec.\ \ref{restba};
the details of the computer simulation of quantum transport are given in Appendix~\ref{nummodmat}. 
The conclusions are given in Sec.\ \ref{conclu}.

\begin{figure*}[t]
\includegraphics[width=0.7\linewidth]{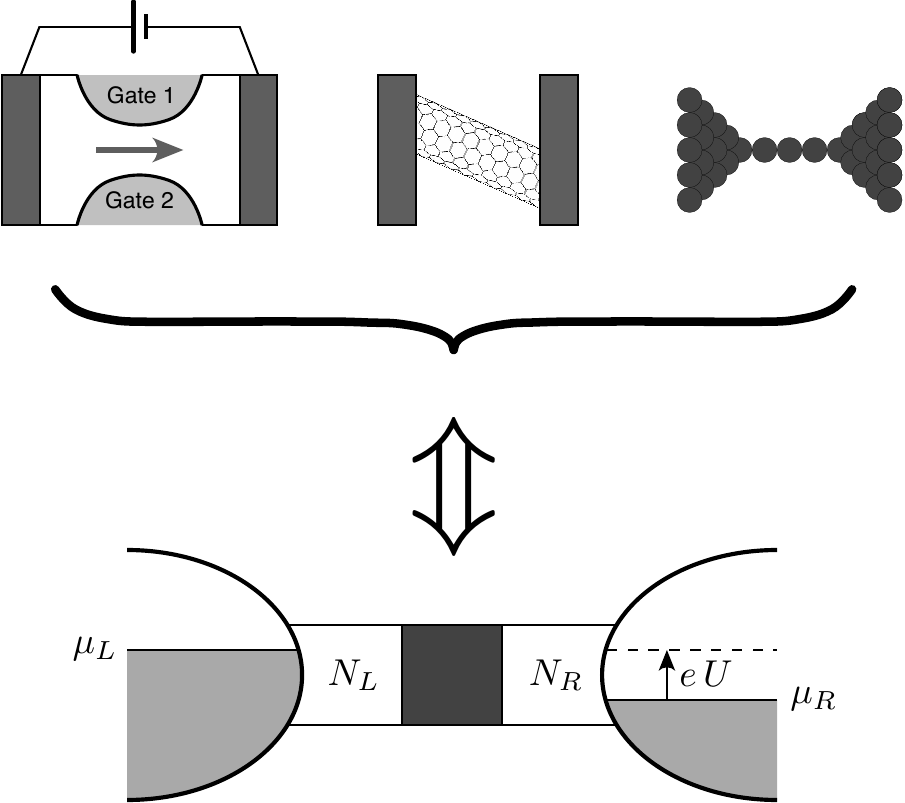}
\caption{ \label{landauer:fig}
Physical suppositions behind the Landauer-B\"{u}ttiker formalism.
Top: Basic nanoscopic systems; from left: a~quantum point contact (QPC) in semiconducting heterostructure, a~carbon nanotube, and monoatomic quantum wire (each system is contacted by two electrodes and connected to a~voltage source driving a~current, as shown for QPC).
Bottom: A~theoretical model, containing the two macroscopic reservoirs (left and right) with fixed chemical potentials ($\mu_L$, $\mu_R$), waveguides with their numbers of normal modes ($N_L$, $N_R$), and the central region (dark square) for which transmission probabilities ($T_n$) need to be determined by solving a~relevant quantum-mechanical wave equation. 
}
\end{figure*}

\section{Landauer-B\"{u}ttiker transport in nanoscopic systems
  and graphene}
\label{lanbutgra}

\subsection{Remark on the origin of zero-temperature Landauer-Sharvin
  resistance}
First, let's look for a concise answer to the question:
{\em Where does electrical resistance come from at absolute zero?}

In the familiar Drude model of electrical conduction \cite{Kit05}
electrons are assumed to constantly bounce between heavier, stationary
lattice ions, allowing one to express the material specific resistivity
as a~function of the electron's effective mass, velocity, and the mean
free path. In quantum-mechanical decription of solids, Drude model
provides a~reasonable approximation as long as the Fermi wavelength remains
much shorter than electron's mean free path and the conductor size.

The picture sketched above changes substantially when electic charge
flows through a~nanoscopic system, such as quantum point contact in
semiconducting heterostructure \cite{Wee88}, a~carbon nanotube \cite{Whi98},
or monoatomic quantum wire \cite{Agr03} (see Fig.\ \ref{landauer:fig},
top part).
Assuming for simplicity that such a system has no internal degrees of freedom
leading to the degeneracy of quantum states, in other words --- that in
a~sufficiently small energy range $\Delta{}E$ we have at most one quantum
state (level) --- we note that the time of flight of an electron through
the system is limited from below by the time-energy uncertainty relation
\begin{equation}
  \Delta{}t\geqslant{}\frac{\hbar}{\Delta{}E}. 
\end{equation}
Next, by linking the energy range $\Delta{}E$ with the electrochemical
potential difference in macroscopic electrodes (reservoirs) connected
to the nanoscopic system (see Fig.\ \ref{landauer:fig}, bottom part),
we can write
\begin{equation}
  \Delta{}E=\mu_L-\mu_R=eU, 
\end{equation}
where $U$ denotes the difference in electrostatic potential on both sides of
the system, and is the elementary charge (without sign). Combining the above
equations we obtain the limit for the electric current flowing through the
system,
\begin{equation}
  I=\frac{e}{\Delta{}t}\leqslant{}\frac{e^2}{\hbar}U,
\end{equation}
which means that the electrical conductivity
\begin{equation}
\label{gubound}
  G=\frac{I}{U}\leqslant{}\frac{e^2}{\hbar}. 
\end{equation}

We thus see that the uncertainty principle of energy and time leads to
a~finite value of the conductivity, and therefore to a~nonzero value of
the electrical resistance, of a nanoscopic system.
By rigorous derivation, the upper bound in Eq.\ (\ref{gubound})
is replaced by $e^2/h$, introducing the Landauer-Sharvin resistance
in noninteracting electron systems \cite{Sha65,Lan57,But85}. 
Obviously, many-body effects may alter this conclussion substantially
\cite{Mei92}. 
For instance, the resistivity of graphene sample may drop below the
Landauer-Sharvin bound due to hydrodynamic effects \cite{Kum22}.
In twisted bilayer graphene, both the interaction-driven insulating and 
superconducting (i.e., resistance-free) phases were observed
\cite{Cao18a,Cao18b,Fid18}. These issues are, however, beyond the scope
of the present work.

\subsection{The Landauer-B\"{u}ttiker formula}
At a temperature close to absolute zero ($T\rightarrow{}0$) and in the limit
of linear response, i.e., the situation in which the electrochemical potential
difference also tends to zero ($\mu_L-\mu_R=eU\rightarrow{}0$), it can be
shown that the electrical conductivity of a nanoscopic system is proportional
to the sum of transition probabilities for the so-called normal modes in the
leads \cite{Naz09}, 
\begin{equation}
\label{gland}
  G=g_0\sum_n{}T_n(E_F), 
\end{equation}
where $g_0$ denotes the conductance quantum; namely, $g_0=2e^2/h$ for systems
exhibiting spin degeneracy (for graphene, we have $g_0=4e^2/h$ due to the
additional degeneracy --- called valley degeneracy --- related to the presence
of two nonequivalent Dirac points in the dispersion relation). 
The probabilities ($T_n$) are calculated by solving (exactly or approximately)
the corresponding wave equation (Schr\"{o}dinger or Dirac) for a~fixed energy,
which, given the assumptions made, can be identified with the Fermi energy
$E_F$. Importantly, we perform the calculations under the additional
assumption that there are so-called waveguides between the macroscopic
reservoirs and the nanoscopic system, for which we can provide (for a~fixed
value of $E_F$) solutions in the form of propagating waves, the number of
which is $N_L$ or $N_R$, for the left or right waveguide, respectively
(see Fig.\ \ref{landauer:fig}).
We also assume that any wave incident on  
a waveguide-reservoir interface, coming from the system, is always
absorbed in the reservoir. (For the discussion of possible alternative
assumptions, see Ref.\ \cite{But85}.)  

It is worth noting that the sum appearing in formula (\ref{gland}) is the 
trace of the transmission matrix, the value of which does not depend on the
choice of the basis; therefore, it can be expected that the result does not
depend on how precisely we construct the aforementioned waveguides, which,
it is worth emphasising, are an auxiliary construction that usually has no
direct physical interpretation. (For the same reason, the result will be
the same whether we consider scattering from left to right or in the opposite
direction.)
In the context of graphene, it was soon shown that for various types of
waveguides, including normal conductors modeled by square lattices, a broad
window of parameters can be identified such that the waveguides appear
transparent, i.e., the transport properties are governed by the central
region \cite{Two06,Bla07,Rob07}. 

As mentioned above, the details of the calculations (or computer simulations)
leading to the determination of the probability values ($T_n$) will depend
on the geometry of the system under consideration.
If waveguides are modelled as strips of fixed width ($W$), at the edges of
which the wave function disappears, the normal modes have the form of plane
waves \cite{planewfoo}, for which the longitudinal component of the wave
vector ($k_x$) is continuous and the normal component ($k_y$) is quantized
according to the rule
\begin{equation}
\label{kynw}
  k_y^{(n)}=\frac{\pi{}n}{W}, \ \ \ \ n=1,2,\dots. 
\end{equation}
The calculations are particularly simple in cases where the central region
(marked with a dark square in Fig.\ \ref{landauer:fig}) differs from the
leads only in that it contains an electrostatic potential that depends on
the $x$ coordinate (oriented along the main axis of the system), for example
in the form of a~rectangular barrier.
Then the transmission matrix has a~diagonal form (no scattering between
normal modes occurs), and in special cases, such as the rectangular barrier
mentioned above, but also e.g.\ the parabolic potential considered by Kemble
in 1935 \cite{Kem35}, it is possible to provide compact analytical formulas.

We will not present the exact results here, but only point out that for
solutions obtained by the mode-matching method for the Schrödinger equation,
one can write approximately
\begin{equation}
\label{tnappx}
  T_n=T(k_y^{(n)})\approx
  \begin{cases}
    1\ \ \text{if }\ k_y\leq{}k_F, \\
    0\ \ \text{if }\ k_y>k_F,
  \end{cases}
\end{equation}
which we write more briefly as $T_n\approx\Theta(k_F-k_y^{(n)})$, with
$\Theta(x)$ denoting the Heaviside step function.
In Eq.\ (\ref{tnappx}) we introduce the wave vector $k_F$ corresponding to
the Fermi energy $E_F$ (assuming for simplicity that the dispersion relation
is isotropic) calculated with respect to the top of the potential barrier in
the central region.
Furthermore, assuming that there are many modes for which $k_y^{(n)}<k_F$
(which occurs if $k_FW\gg{}1$), and therefore the summation in Eq.\ 
(\ref{gland}) can be replaced to a good approximation by integration,
we obtain --- via Eqs.\ (\ref{kynw}) and (\ref{tnappx}) --- the result known
in the literature as the Sharvin conductance \cite{Sha65}
\begin{equation}
\label{gshar}
  G_{\rm Sharvin}\approx{}g_0\frac{k_FW}{\pi}. 
\end{equation}

It is worth noting that the reasoning leading to Eq.\ (\ref{gshar}) can
be relatively easily applied to the case where the electrostatic potential
in the central region is approximately constant and the width of the
conducting region is a function of the position along the longitudinal axis
($x$), changing slowly enough that the scattering between normal modes can
be neglected. The above-mentioned case is the so-called quantum point
contact (QPC), shown schematically in Fig.\ \ref{landauer:fig} (top part),
which can be realized in semiconductor heterostructures hosting
a~two-dimensional electron gas (2DEG) \cite{Naz09}.

In contrast to bulk systems, for which the semiclassical Drude-Boltzmann
approach works relatively well \cite{Das11,Kos14}, nanosystems with spatially
confined electrons exhibit several quantum effects that can be much better
grasped within the Landauer-B\"{u}ttiker formalism. 
Additionally, transport in graphene
at low carrier concentration is governed by evenescent modes, inclusion
of which in Drude-Boltzmann description is rather problematic \cite{Ryc21a}.
For instance, Yoshino and Murata \cite{Yos15} assumed linear relaxation time
on energy dependence, leading to nonzero conductivity at the charge-neutrality
point, but the physical reasoning behind such an assumption seems unclear. 
On the other hand, the Landauer-B\"{u}ttiker formalism naturally includes both
evanescent and propagating solutions, allowing to describe the phenomena
such as the universal minimal conductivity of monolayer graphene.
(Further details are given in Sec.\ \ref{secunisig}.)

\subsection{Shot noise and counting statistics}
The second quantity, besides electrical conductivity, that characterizes
nanoscopic systems at temperatures close to absolute zero is the shot-noise
power. For the sake of brevity, let us point out the basic facts:
First, the electric charge $Q$ flowing through the system shown schematically
in Fig.\ \ref{landauer:fig} (lower part) in a short time interval $\Delta{}t$
is a~random variable.
Second, the expectation value of such a~variable is closely related to
the electrical conductivity $G$ in the linear-response limit,
\begin{equation}
\label{averqgut}
  \langle{}Q\rangle=GU\Delta{}t \ \ \ \ (U\rightarrow{}0). 
\end{equation}
The reason the measured value of $Q$ fluctuates at successive time intervals
is due to the discrete (granular) nature of the electric charge. 

Assuming (for the moment) that electrons jump from one reservoir to another
completely independently, we conclude that the charge flow is a Poisson
process, or more precisely, that the quantity $Q/e$ follows the Poisson
distribution; the variance is therefore proportional to the expectation
value given by Eq.\ (\ref{averqgut}),
\begin{equation}
  \label{varqpoi}
  \left\langle{}Q^2-\langle{}Q\rangle^2\right\rangle_{\rm Poisson}=
  e\langle{}Q\rangle=eU\Delta{}tg_0\sum_n{}T_n. 
\end{equation}
More generally, $m$-th central moment can be written as
\begin{equation}
\label{mmomqpoi}
  \langle\langle{}Q^m\rangle\rangle_{\rm Poisson} \equiv
  \left\langle{}\left(Q-\langle{}Q\rangle\right)^m\right\rangle_{\rm Poisson}
  = e^{m-1}\langle{}Q\rangle, 
\end{equation}
with the integer $m\geqslant{}1$.

The Fano factor, quantifying the shot-noise power, is defined as the ratio
of the actual measured variance of the charge flowing through the system to
the variance given by Eq.\ (\ref{varqpoi}), or more precisely
\begin{equation}
\label{fanodef}
  F=\frac{\left\langle{}Q^2-\langle{}Q\rangle^2\right\rangle}{%
  \left\langle{}Q^2-\langle{}Q\rangle^2\right\rangle_{\rm Poisson}}=
  1-\frac{\sum_n{}T_n^2}{\sum_n{}T_n}.
\end{equation}
(For compact derivation, see e.g.\ Ref.\ \cite{Naz09}.)
In the following, we have limited our considerations to long time intervals
such that $eU\Delta{}t\gg{}\hbar$; hence $F$ characterizes the zero-frequency
noise, not to be confused with the celebrated $1/f$ noise in electronic
systems \cite{Tak93}. 
A~generalization of Eq.\ (\ref{fanodef}) for finite
times (and nonzero temperatures) is also possible \cite{Sch07}. 

In particular, it follows from Eq.\ (\ref{fanodef}) that the Poisson limit
($F\rightarrow{}1$) is realized in the case of a~tunnel junction, for which
we have $T_n\ll{}1$ for each $n$. This is a completely different case than
the ballistic system considered above, which exhibits Sharvin conductance;
then, replacing the summation with integration as before and using the
approximation given by Eq.\ (\ref{tnappx}), we obtain
\begin{equation}
\label{ffsharvin}
  F_{\rm Sharvin}\approx{}
  1 - \frac{\int{}dk_y\left({\Theta}(k_F-k_y)\right)^2}{%
  \int{}dk_y{}\,{\Theta}(k_F-k_y)}
  \approx{}0.
\end{equation}
In general, for fermionic systems we always have $0<F<1$; the factor $1-T_n$
appearing in the numerator in Eq.\ (\ref{fanodef}) is a~consequence of
the Pauli exclusion principle. In the case of the idealized ballistic system
we have $F=0$, see Eq.\ (\ref{ffsharvin}), which means that the electron
count ($Q/e$) does not fluctuate with time.
One could say that the electrons avoid each other so much that they "march"
at equal intervals. 
(Of course, this is only possible at absolute zero temperature, otherwise
additional thermal noise appears, i.e. the Nyquist-Johnson noise proportional
to the conductivity value, whose influence we have ignored here;
see Ref.\ \cite{Naz09}.)

In an attempt to determine higher charge cumulants, it is convenient to
introduce characteristic function 
\begin{equation}
  \Lambda(\chi)=\left\langle\,\exp(i\chi{Q}/e)\,\right\rangle,
\end{equation}
such that
\begin{equation}
\label{mmomqlam}
   \langle\langle{}Q^m\rangle\rangle \equiv
   \left\langle{}\left(Q-\langle{}Q\rangle\right)^m\right\rangle =
   e^m\left.
   \frac{\partial^m{}\ln\Lambda(\chi)}{\partial(i\chi)^m}\right|_{\chi=0}. 
\end{equation}
Assuming $U>0$ for simplicity, we arrive at the Levitov-Lesovik formula
\cite{Naz09,Sch07}
\begin{equation}
\label{lamlele}
  \ln\Lambda(\chi)=\frac{g_0U\Delta{t}}{e}\sum_n
  \ln\left[1+T_n\left(e^{i\chi}-1\right)\right], 
\end{equation}
expressing the full counting for noninteracting fermions. 

Substitution of the above into Eq.\ (\ref{mmomqlam}) with 
$m=1$ and $m=2$ reproduces (respectively) Eqs.\ (\ref{gland})
and (\ref{fanodef}). Analogously, for $m=3$ and $m=4$, we get
\begin{widetext}
\begin{align}
  R_3 &\equiv \frac{\langle\langle{Q^3}\rangle\rangle}{%
  \langle\langle{Q^3}\rangle\rangle_{\rm Poisson}}=
  \left(\sum_n{}T_n-3\,\sum_n{}T_n^2+2\,\sum_n{}T_n^3\right)\Big/
  \sum_n{}T_n,
  \label{r3def} \\
  R_4 &\equiv \frac{\langle\langle{Q^4}\rangle\rangle}{%
  \langle\langle{Q^4}\rangle\rangle_{\rm Poisson}}=
  \left(\sum_n{}T_n-7\,\sum_n{}T_n^2+12\,\sum_n{}T_n^3
  -6\,\sum_n{}T_n^4\right)\Big/\sum_n{}T_n.
  \label{r4def}
\end{align}
\end{widetext}
For the Sharvin regime, see Eq.\ (\ref{tnappx}),
\begin{equation}
\label{r34sharvin}
  \left(R_3\right)_{\rm Sharvin} \approx \left(R_4\right)_{\rm Sharvin} \approx 0. 
\end{equation}

\begin{figure*}[t]
\includegraphics[width=\linewidth]{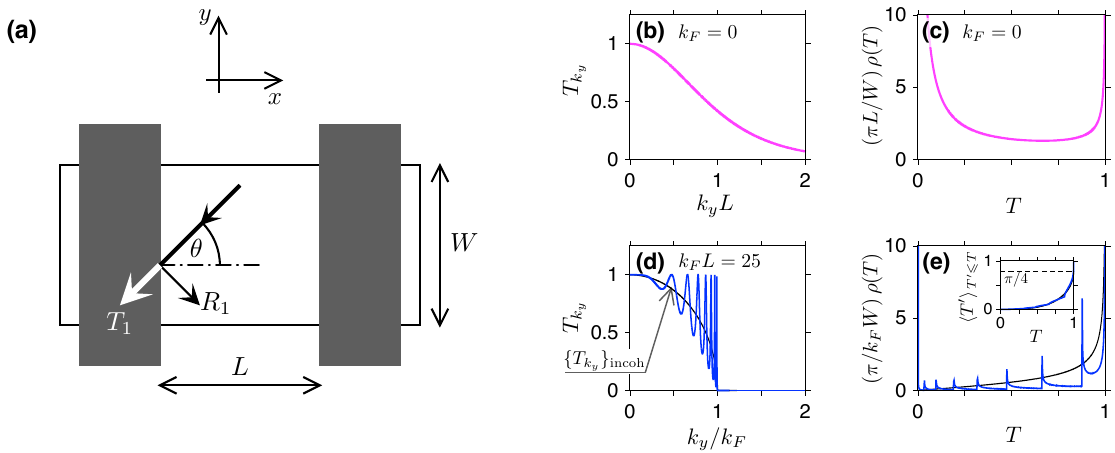}
\caption{ \label{stripttky}
  (a) Rectangular graphene sample (white area) of the width $W$ contacted
  to the leads (dark areas) at a~distance $L$. The coordinate system $(x,y)$
  is also shown. Scattering of Dirac electrons
  at a~sample-lead interface for the incident angle $\theta$ is characterized
  by the transmission ($T_1$) and the reflection ($R_1$) coefficients given
  by Eq.\ (\ref{t1r1costh}).
  (b) Transmission probability for a~double barrier [see Eq.\ (\ref{tt12phi})]
  as a~function of the transverse momentum $k_y$ and
  (c) the corresponding distribution of transmission probabilities at the Dirac
  point $k_F=0$ (with $k_F=|E|/\hbar{}v_F$).
  (d,e) Same as (b,c) but the doping fixed at $k_FL=25$. Blue lines
  represent the exact results, black lines depict the approximation
  $\{T_{k_y}\}_{\rm incoh}$ given by Eq.\ (\ref{ttkyincoh}).
  Inset in (e) shows the integrated distribution
  $\langle{}T'\rangle_{T'\leqslant{}T}=(\pi/k_FW)\int_0^{T}dT'T'\rho(T')$ for both
  the exact $\rho(T)$ [blue line] and the approximation given by Eq.\
  (\ref{rhottapp}) [black line].
  The sub-Sharvin value of $\langle{}T\rangle=\pi/4$ is depicted with dashed
  horizontal line. 
}
\end{figure*}

\subsection{Scattering of Dirac fermions in two dimensions}
Using the introductory information gathered above, we will now calculate
--- with some additional simplifying assumptions --- the electrical
conductivity as well as the higher charge cimulants of a~graphene strip.
The effective wave equation for itinerant electrons in this two-dimensional
crystal is the Dirac-Weyl equation, the detailed derivation of which can be
found, e.g., in Katsnelson's textbook \cite{Kat20}, and which can be written
in the form
\begin{equation}
\label{diraceq}
  \left[v_F\,{\boldsymbol p}\cdot{\boldsymbol\sigma}+V(x)\right]\Psi=E\Psi,
\end{equation}
where the energy-independent Fermi velocity is given by
$v_F=\sqrt{3}t_0a/(2\hbar)$, where $t_0\approx{}2.7\,{\rm eV}$ denotes the
nearest-neighbor hopping integral in the graphene plane and
$a=0.246\,{\rm nm}$ is the lattice constant (as a result, the approximate
value of $v_F$ is about $10^6\,{\rm m/s}$, which is several times lower
than typical Fermi velocities in metals).
The remaining symbols in Eq.\ (\ref{diraceq}) are the quantum mechanical
momentum operator ${\boldsymbol p}=-i\hbar\left(\partial_x,\partial_y\right)$
(the notation $\partial_j$ here means differentiation with respect to the
selected coordinate, $j=x,y$),
${\boldsymbol \sigma}=\left(\sigma_x,\sigma_y\right)$ is a vector composed
of Pauli matrices \cite{valleyfoo}, and the electrostatic potential energy
$V(x)$ is assumed to depend only on the position along the principal axis of
the system.

The above assumptions imply that we can look for solutions to
Eq.\ (\ref{diraceq}) in the form of a~two-component (i.e., spinor) wave
function
\begin{equation}
  \Psi=\begin{pmatrix}
           \phi_a \\
           \phi_b
         \end{pmatrix}
  e^{ik_y{}y},
\end{equation}
where $\phi_a$  and $\phi_b$ are functions of $x$.
By substituting the above ansatz into Eq.\ (\ref{diraceq}) we obtain a~system
of ordinary differential equations
\begin{align}
  \phi_a' &= k_y\phi_a+i\frac{E-V(x)}{\hbar{}v_F}\phi_b,
  \label{phiaprim} \\
  \phi_b' &= i\frac{E-V(x)}{\hbar{}v_F}\phi_a-k_y\phi_b,
  \label{phibprim}
\end{align}
where the primes on the left-hand side denote derivatives with respect to $x$. 
We see that in the system of Eqs.\ (\ref{phiaprim}), (\ref{phibprim}) the
quantities $k_y$ and $E$ play the role of parameters on which the solutions
depend (in the following, when calculating, among others, the electrical
conductivity, we will identify the electron's energy with the Fermi energy
by setting $E=E_F$).

At this point it is worth to comment on the problem of quantizing the value of
the transverse momentum ($k_y$) in Eqs.\ (\ref{phiaprim}), (\ref{phibprim}).
Assuming that the component of the current density perpendicular to the axis
of the graphene strip disappears at its edges (i.e., for $y=0$ and $y=W$, 
see Fig.\ \ref{stripttky}(a)),
what is known as the so-called mass confinement \cite{Ber87}, we get
a~slightly different quantization than in the case of the Schr\"{o}dinger
system, see Eq.\ (\ref{kynw}), namely
\begin{equation}
\label{kynpiw} 
  k_y^{(n)}=\frac{\pi{}(n+1/2)}{W}, \ \ \ \ n=0,1,2,\dots. 
\end{equation}
In practice, however, the assumptions made in the following part mean that
when calculating measurable quantities ($G$, $F$, etc.) we will approximate
the sums appearing in Eqs.\ (\ref{gland}), (\ref{fanodef}), (\ref{r3def}),
(\ref{r4def}), with integrals with respect to $dk_y$; the quantization change
described above is therefore insignificant for further considerations.

The solution of the system of Eqs.\ (\ref{phiaprim}), (\ref{phibprim})
is particularly simple in the case if the electrostatic potential energy,
i.e. the function $V(x)$, is piecewise constant.
Then, the solutions in individual sections (i.e., areas where $V(x)$ is
constant) have the form of plane waves.
For instance, for $E>V(x)$ waves traveling in the positive $(+)$ and negative
$(-)$ directions along the $x$ axis, look as follows
\begin{equation}
\label{phipm}
\phi^{(+)}=\begin{pmatrix}
           1 \\
           e^{i\theta}
         \end{pmatrix}
e^{ik_x{}x},
\ \ \ \
\phi^{(-)}=\begin{pmatrix}
           1 \\
           -e^{-i\theta}
         \end{pmatrix}
e^{-ik_x{}x},
\end{equation}
where we have defined
\begin{align}
  e^{i\theta}&=(k_x+ik_y)/k_F, \nonumber \\
  k_F&=\left(E-V(x)\right)/\hbar{}v_F, \label{kfdef} \\ 
  \text{and }\ \ \ k_x&=\sqrt{k_F^2-k_y^2}. \nonumber
\end{align}
For $E<V(x)$, propagating-wave solutions also exist (this is, by the way,
the main difference between the solutions of the massless Dirac equation
and the Schr\"{o}dinger equation, which leads in particular to
the phenomenon known as Klein tunneling \cite{Rob12,Gut16}) and differ from
those given in Eq.\ \ref{phipm} only in some signs. 
We leave the straightforward derivation to the reader. 

At the interface of regions differing in the (locally constant) value of
$V(x)$, we perform a~matching of wave functions, which for the two-dimensional
Dirac equation reduces to solving the continuity conditions for both spinor
components \cite{matchfoo}. For instance, if we consider the scattering from
the right side of the discontinuity to the left side, we write
\begin{equation}
\label{trphilphir}
  t\phi^{(L,-)}=\phi^{(R,-)}+r\phi^{(R,+)},
\end{equation}
where the spinor functions with indices $L$ and $R$ differ in the values of
$k_F$ and $k_x$ [see Eqs.\ (\ref{phipm}), (\ref{kfdef})], but are
characterized by the same value of $k_y$.
Since the considerations concern the interface between the graphene sample
and the graphene region covered with a~metal electrode (see Fig.\
\ref{stripttky}(a)), the calculations can be simplified by adopting the model
of a~heavily doped electrode, in which we set $V(x)=-V_\infty$, where
$V_\infty\rightarrow\infty$; we can then write the wave functions on the left
in asymptotic form
\begin{equation}
  \phi^{(L,\pm)}\simeq\begin{pmatrix}
           1 \\
           \pm{}1
  \end{pmatrix},
\end{equation}
where we have omitted the phase factor, which is not important for further
considerations. After substituting the above into Eq.\ (\ref{trphilphir}),
the calculations are straightforward; we now present the results for
the transition and reflection probabilities
\begin{equation}
\label{t1r1costh}
  T_1=|t|^2=\frac{2\cos\theta}{1+\cos\theta},
  \ \ \ \
  R_1=|t|^2=\frac{1-\cos\theta}{1+\cos\theta}, 
\end{equation}
which turn out to depend only on the angle of incidence $\theta$ of the plane
wave, or --- more precisely --- on the value of
$\cos\theta=\sqrt{1-(k_y/k_F)^2}$. In particular, we see that for $\theta=0$
we have $T_1=1$ (and $R_1=0$), which is a manifestation of the Klein tunneling
mentioned above (let us emphasize that the potential barrier considered here
has an infinite height). 

The probability of passing through the entire graphene sample, i.e.,
through two electrostatic potential steps occurring at the sample-lead
interface (see Fig.\ \ref{stripttky}(a)), is most easily calculated using
the double-barrier formula, the clear derivation of which can be found, e.g.,
in the Datta's handbook \cite{Dat97}
\begin{equation}
\label{tt12phi}
  T_{12}=\frac{T_1{}T_2}{1+R_1R_2-2\sqrt{R_1R_2}\cos\phi},
\end{equation}
where a~phase shift
\begin{equation}
  \phi=k_xL=L\sqrt{k_F^2-k_y^2}, 
\end{equation}
related to the propagation of a plane wave along the main axis $x$, is 
introduced. (Note here that the phase shift introduced in this manner also implies the assumption that any reflections from the side edges of the system do not change the value of $k_y$; in practice, this implies that we restrict our considerations to systems for which $W\gg{}L$.)
Assuming barrier symmetry, $T_2=T_1$, $R_2=R_1$, and substituting the formulas
given in Eq.\ (\ref{t1r1costh}), we can now write $T_{12}$ explicitly as
a~function of $k_y$ and $E$,
\begin{equation}
\label{t12tky}
  T_{12}=T_{k_y}(E) = \left[ 
    1+\left(\dfrac{k_y}{\varkappa}\right)^2\sin^2\left(\varkappa{}\,L\right)
  \right]^{-1}, 
\end{equation}
where 
\begin{equation}
\label{varkapp}
\varkappa = \begin{cases}
  \sqrt{k_F^2-k_y^2}, & \text{for }\  |k_y|\leqslant{}k_F, \\
  i\sqrt{k_y^2-k_F^2}, & \text{for }\  |k_y|>k_F, \\
  \end{cases}
\end{equation}
and the Fermi wave vector, assuming $V(x)=0$ for the sample region,
is equal to $k_F=|E|/(\hbar{}v_F)$. The absolute value in the last expression
arises from the fact that formulas in Eq.\ (\ref{t1r1costh}) and the following
results are identical for $E<0$; we leave the verification of this
property to the reader.

\subsection{The conductivity, shot noise, and higher cumulants
for ballistic graphene strip}
\label{secunisig} 

The physical consequences of the above expression for the transition
probabilities $T_{k_y}(E)$, see Eqs.\ (\ref{t12tky}) and (\ref{varkapp}),
are now discussed for two physical situations: a~charge-neutral sample
($k_F=0$) and the Sharvin limit ($k_FW\gg{}1$). 
(Unless otherwise stated, we also assume geometry with long, parallel
sample-lead interfaces and $W\gg{}L$.)  

In the first case ($k_F=0$) we obtain $\varkappa=i|k_y|$ and can use the
identity $\sin(ix)=i\sinh x$, resulting in a~surprisingly simple expression
\begin{equation}
\label{tky0}
  T_{k_y}(0)=\frac{1}{\cosh^2(k_yL)},  
\end{equation}
visualized in Fig.\ \ref{stripttky}(b). 
In the wide-sample limit, $W\gg{}L$, the sums appearing in the formulas
for Landauer conductance $G$ [see Eq.\ (\ref{gland})], Fano factor $F$ 
[Eq.\ (\ref{fanodef})], and higher cumulants $R_3$, $R_4$
[Eqs.\ (\ref{r3def}) and (\ref{r4def})] can be approximated with integrals
[see also Eq.\ (\ref{kynpiw})], leading to
\begin{align}
  \sum_n{}T_n(0) &\approx{}
  \frac{W}{\pi}\int_0^{\infty}dk_y\,T_{k_y}(0)=
  \frac{W}{\pi{}L}, \\
  \sum_n{}\left[T_n(0)\right]^2 &\approx{}
  \frac{W}{\pi}\int_0^{\infty}dk_y\,\left(T_{k_y}(0)\right)^2=
  \frac{2}{3}\frac{W}{\pi{}L}, 
\end{align}
or, more generally,
\begin{equation}
  \label{sumpdiff}
  \sum_n{}\left[T_n(0)\right]^m \approx{}
  \frac{W}{2\sqrt{\pi}{}\,L}\,
  \frac{\Gamma(m)}{\Gamma(m+1/2)}\ \ \ \ \text{for }\ m>0,  
\end{equation}
where $\Gamma(x)$ is the Euler gamma function.
To facilitate future comparisons with other transport regimes, we will
additionally define 
\begin{equation}
\label{avertmdiff}
  \langle{}T^m\rangle_{k_F=0}= L\int_0^{\infty}dk_y\,\left(T_{k_y}(0)\right)^m
  =\frac{\sqrt{\pi}\,\Gamma(m)}{2\Gamma(m+\frac{1}{2})},
\end{equation}
such that $\langle{}T\rangle_{k_F=0}=1$. 
Taking into account the
graphene-specific fourfold degeneracy of states due to the presence of spin
and valley degrees of freedom (the conductance quantum is therefore
$g_0=4e^2/h$), we obtain
\begin{gather}
  G\approx\frac{4e^2}{\pi{}h}\frac{W}{L},\ \ \ \
  F\approx1-\frac{2}{3}=\frac{1}{3},
  \label{gfpdiff} \\
  R_3 \approx \frac{1}{15},\ \ \ \ R_4\approx{}-\frac{5}{512}.
  \label{r34pdiff} 
\end{gather}

The value of $G\propto{}W/L$ (instead of $G\propto{}W$, as in a~typical
ballistic system) means that charge-neutral graphene exhibits universal
specific conductivity, $\sigma_0=4e^2/(\pi{}h)$, the value of which is
additionally determined only by the universal constants of nature.
The value of the Fano factor $F=1/3$ is also not accidental, as it is
a~value characteristic for ohmic (disordered) conductors.
(The same applies to higher cumulants.) 
In the context of graphene, the term {\em pseudodiffusive conductivity}
is often used to emphasize that this ballistic system perfectly emulates
an ohmic conductor within the appropriate parameter range.
It should be emphasized that the first two theoretical values,
given in Eq.\ (\ref{gfpdiff}) and originally derived in Refs.\ \cite{Kat06}
(conductance) and \cite{Two06} (Fano factor), have been experimentally
confirmed with satisfactory accuracy in 2008 \cite{Dan08}.
(For the comprehensive theoretical discussion of full counting statistics
for graphene at the Dirac point, see Ref.\ \cite{Sch10}.)

\begin{figure*}[t]
\includegraphics[width=\linewidth]{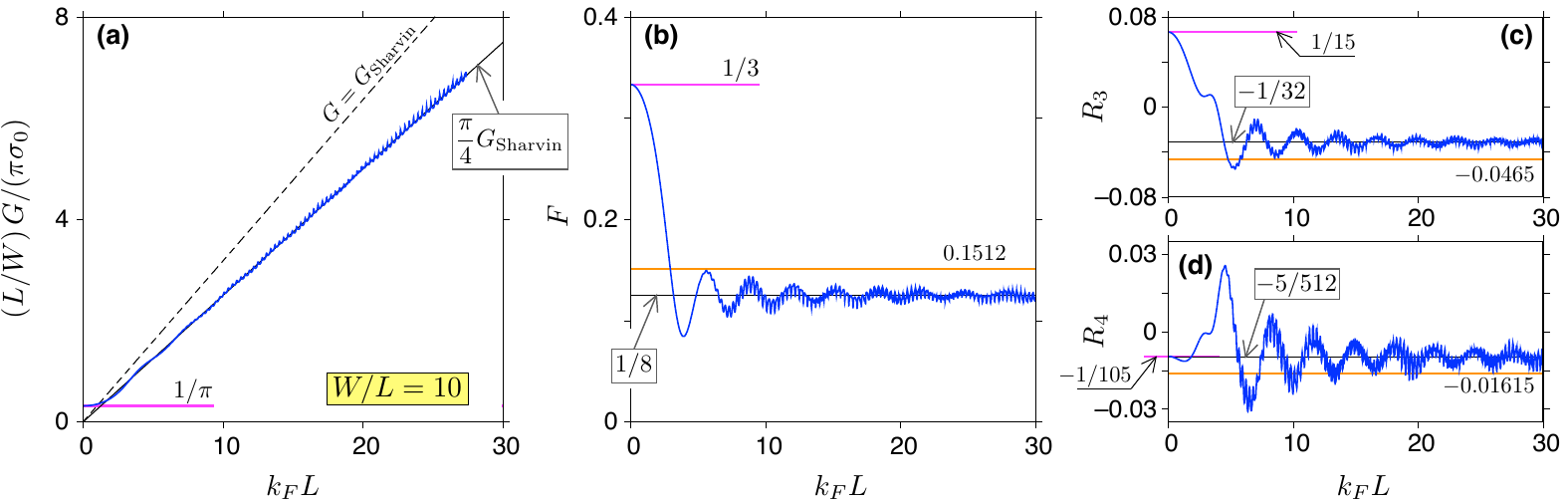}
\caption{ \label{gfr34strip}
  Conductance (a), Fano factor (b), third (c), and fourth (d) charge-transfer
  cumulant for graphene strip dislayed as functions of the Fermi momentum
  (solid blue lines).
  The aspect ratio is fixed at $W/L=10$.
  Dashed black line in (a) depicts the Sharvin conductance
  $G_{\rm Sharvin}=g_0k_FW/\pi$, with $g_0=4e^2/h$; the sub-Sharvin values,
  given by Eqs.\ (\ref{gsubshar}), (\ref{fr34subshar}) are depicted with
  solid black lines in all panels.
  Short purple line (a--d) marks the pseudodiffufive value, see Eqs.\
  (\ref{gfpdiff}), (\ref{r34pdiff}), approached for $k_F\rightarrow{}0$. 
  Wide orange lines (b--d) depict the values following from the approximated
  distribution of transmission probabilities $\rho_{\rm approx}(T)$, see Eq.\
  (\ref{rhottapp}).
}
\end{figure*}

\begin{figure*}[t]
\includegraphics[width=\linewidth]{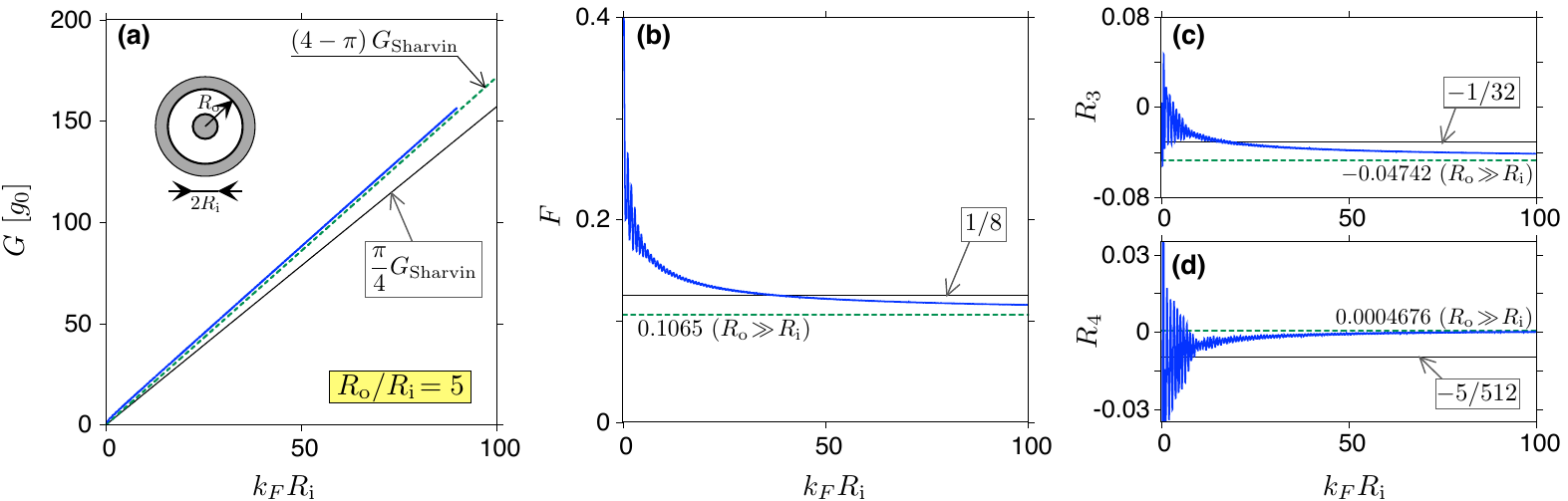}
\caption{ \label{gfr34corb}
  (a--d) Same as Fig.\ \ref{gfr34strip} but for the Corbino disk, see inset
  in (a), with the outer-to-inner radii ratio $R_{\rm o}/R_{\rm i}=5$.
  Solid blue lines mark the exact results following from Eqs.\
  (\ref{tjhank}), (\ref{ddnupm}). 
  Remaining lines mark the sub-Sharvin values relevant for the thin-disk
  limit $R_{\rm i}/R_{\rm o}\rightarrow{}1$ [solid black] and for the
  narrow-opening limit, $R_{\rm i}/R_{\rm o}\rightarrow{}0$, see Eq.\
  (\ref{gfr34narrow}) [dashed green]. 
}
\end{figure*}

In the Sharvin limit ($k_FW\gg{}1$) the situation looks a~bit different.
We can then assume that the contribution of modes for which $k_y>k_F$
(i.e., {\em evanescent} modes) is negligible and limit the considerations
to $k_y\leqslant{}k_F$.
Next, we notice that as the values of $T_{k_y}(E)$ (or their powers) are summed,
the $\sin^2\left(L\sqrt{k_F^2-k_y^2}\right)$ term of Eq.\ (\ref{t12tky}) 
oscillates rapidly, especially as $k_y$ approaches $k_F$.
Therefore, it seems reasonable to replace the sine argument with a~random
phase and average the result, which leads to the following approximation
\begin{align}
  T_{k_y}(E) \approx
  \{T_{k_y}\}_{\rm incoh} &= \frac{1}{\pi}
  \int_0^{\pi}\frac{d\varphi}{1+\left(k_y^2/\varkappa^2\right)
    \sin^2\varphi} \nonumber\\
    &=
  \sqrt{1-\left(k_y/k_F\right)^2}, \label{ttkyincoh}
\end{align}
where we used the table integral \cite{Ryz-2-553-3}
\begin{equation}
\label{intabu}
  I(a,b) = \frac{1}{2\pi}\int_{-\pi}^{\pi}\frac{du}{a+b\cos{}u}=
  \frac{1}{\sqrt{a^2-b^2}}, \ \ \ \ \text{for}\ a>|b|,   
\end{equation}
substituting
\begin{align}
  u &= 2\varphi, \nonumber \\
  a &= \frac{1-\frac{1}{2}\eta^2}{1-\eta^2}, \ \ \ \
  b = a-1 = \frac{\frac{1}{2}\eta^2}{1-\eta^2},
  \label{uabeta} \\
  \text{with }\ \eta &= k_y/k_F. \nonumber
\end{align}
The comparison between the approximation given in Eq.\ (\ref{ttkyincoh})
and the actual $T_{k_y}(E)$, see Eqs.\ (\ref{t12tky}), (\ref{varkapp}),
for $k_FL=25$,
%and $W/L=10$,
is presented in Fig.\ \ref{stripttky}(d). 

Eq.\ (\ref{ttkyincoh}) is essentially the Dirac version of Eq.\ (\ref{tnappx})
describing a~standard ballistic system; when calculating the Landauer
conductance,
we can again approximate the summation by integration and obtain
\begin{equation}
\label{gsubshar}
  G\approx{}\frac{g_0W}{\pi}\int_0^{k_F}
  dk_y\,\{T_{k_y}\}_{\rm incoh}=
  \frac{\pi}{4}G_{\rm Sharvin}, 
\end{equation}
where we recall the value of Sharvin conductance given in Eq.\ (\ref{gshar}). 
The prefactor $\pi/4$ is a~consequence of the fact that in the last expression
in Eq.\ (\ref{ttkyincoh}), where previously there was a step function
$\Theta(k_F-k_y)$, a~term describing an arc of a~circle has appeared;
the conductance of graphene beyond the charge-neutrality point is therefore
reduced compared to a typical ballistic system. 

Interestingly, in deriving Eq.\ (\ref{gsubshar}) we did not explicitly assume,
as in Eq.\ (\ref{gfpdiff}), that the width to length ratio of the sample is
$W/L\gg{}1$; hypothetically, the result given in Eq.\ (\ref{gsubshar}) can
therefore be applied whenever the condition $k_FW\gg{}1$ is satisfied,
regardless of the value of $W/L$. In practice, however, it is difficult to
imagine that the double-barrier transmission formula, Eq.\ (\ref{tt12phi}),
which is the basis of the entire reasoning, could be applied to samples that
do not satisfy the $W/L\gg{}1$ condition. It seems that in the case of
graphene samples with $L\gtrsim{}W$, hard-to-control edge effects can
significantly alter the conductivity \cite{Mia07}.
We will address this issue later in this paper, but first let us calculate
the approximate values of the Fano factor and higher cumulants in the Sharvin
limit.

In order to construct the approximation analogous to this in Eq.\
(\ref{ttkyincoh}), but for $T_{k_y}^2$, it is sufficient to calculate
\begin{align}
\{T_{k_y}^2\}_{\rm incoh} &= \frac{1}{\pi}
\int_0^{\pi}\frac{d\varphi}{\left[1+\left(k_y^2/\varkappa^2\right)
  \sin^2\varphi\right]^2} \nonumber\\
  &=
\sqrt{1-\eta^2}
  \left[1-\frac{1}{2}\eta^2\right], \label{tt2kyincoh}
  \end{align}
where we used the first derivative of $I(a,b)$, see Eqs.\ (\ref{intabu})
and (\ref{uabeta}), with respect to $a$.
More generally, the $m$-th power of $T_{k_y}$ can be approximated by
\begin{widetext}
\begin{align} 
\{T_{k_y}^m\}_{\rm incoh} &= \frac{1}{\pi}
\int_0^{\pi}\frac{d\varphi}{\left[1+\left(k_y^2/\varkappa^2\right)
  \sin^2\varphi\right]^m} 
  = \frac{(-1)^{m-1}}{(m-1)!}\frac{\partial^{m-1}}{\partial{}a^{m-1}}\,I(a,b)
  \nonumber\\
  &=
%\frac{(-1)^{m-1}2^ma^m}{(m-1)!\,(a^2-b^2)^{m+1/2}}\left(\frac{1}{2}-m\right)_m
  \frac{2^{m-1}\Gamma(m-\frac{1}{2})a^{m-1}}{
    \sqrt{\pi}\,\Gamma(m)(a^2-b^2)^{m-1/2}}
  \,{}_2F_1\left(
    1-\frac{m}{2},\frac{1-m}{2};\frac{3}{2}-m;1-\frac{b^2}{a^2}
  \right), \label{tmincoh} \\
  &=   \frac{2^{m-1}\Gamma(m-\frac{1}{2})}{
    \sqrt{\pi}\,\Gamma(m)}\sqrt{1-\eta^2}\,z^{-m+1}
    \,{}_2F_1\left(
      1-\frac{m}{2},\frac{1-m}{2};\frac{3}{2}-m;z
  \right),\ \ \ \ \text{with }\ z=\frac{1-\eta^2}{(1-\frac{1}{2}\eta^2)^2}, 
  \nonumber 
\end{align}
\end{widetext}
where $_2F_1(\alpha,\beta;\gamma;z)$ is the hypergeometric function
\cite{Abr65}. 
(Notice that for a~positive integer $m$, $\alpha=1-\frac{m}{2}$ or
$\beta=\frac{1-m}{2}$ is a~non-positive integer,
so the function reduces to a~polynomial of $z$; after multiplying by
$z^{-m+1}$, the resulting expression can be further simplified to a~degree
$2m-2$ polynomial of the variable $\eta$.) 
For $m=1$ and $m=2$, the above reproduces the results of Eqs.\
(\ref{ttkyincoh}) and (\ref{tt2kyincoh}), respectively.
The next two expressions are
\begin{align}
  \{T_{k_y}^3\}_{\rm incoh} &= \sqrt{1-\eta^2}
  \left(1-\eta^2+\frac{3}{8}\eta^4\right), \\
  \{T_{k_y}^4\}_{\rm incoh} &= \sqrt{1-\eta^2}
  \left(1-\frac{\eta^2}{2}\right)
  \left(1-\eta^2+\frac{5}{8}\eta^4\right). 
\end{align}
Setting $[T_{k_y}(E)]^m\approx{}\{T_{k_y}^m\}_{\rm incoh}$ for
$m=1,\dots,4$, and 
approximating the summations occuring in Eqs.\ (\ref{fanodef}),
(\ref{r3def}), and (\ref{r4def}) by integrations, namely,
\begin{align}
  \sum_n{}\left[T_n(E)\right]^m &\approx{}
  \frac{k_FW}{\pi}\left\langle\{T_{k_y}^m\}_{\rm incoh}\right\rangle 
  \\
  \text{with }\
  \left\langle\{T_{k_y}^m\}_{\rm incoh}\right\rangle
  &= \int_0^{1}d\eta\,\{T_{k_y}^m\}_{\rm incoh}, \label{avertmincoh}
\end{align}
we obtain
\begin{equation}
\label{fr34subshar}
  F \approx{} \frac{1}{8}, \ \ \ \ 
  R_3 \approx{} -\frac{1}{23}, \ \ \ \
  R_4 \approx{} -\frac{5}{512}. 
\end{equation}
(For the first four numerical values of $\langle\{T_{k_y}^m\}_{\rm incoh}\rangle$,
see Table~\ref{avtmtab}.) 

The surprising (non-zero) value of the shot noise in Eq.\ (\ref{fr34subshar})
is close to the experimental results obtained by Danneau {\em et.\ al.\/}
\cite{Dan08}, which are in the range of $F=0.10\div 0.15$.
(The aspect ratio of the sample used in this experiment was $W/L=24$.)
Recently, for other 2D Dirac systems in the gapless limit, a~theoretical
prediction of $F\approx{}0.179$ has been reported \cite{Cha23}. 
It should be emphasized that measuring the shot noise of nanoscopic devices containing components of different materials is rather challenging; 
there are also results in the literature that suggest that the dependence of
the shot noise on the system filling is weak, with the value always close
to the pseudodiffusive $F=1/3$ \cite{Dic08}. 
Unlike for carbon nanotubes, where inelastic processes may lead to
super-Poissonian noise with $1<F<3$ \cite{Ona06}, values of $F>1$
has not been detected for bulk graphene. 
Measurements of higher charge cumulants are so far missing.

In Fig.\ \ref{gfr34strip}, the approximations for $G$, $F$, $R_3$, and $R_4$,
both for $k_F=0$ and for $k_FW\gg{}1$, are compared with
and the actual values following from Eqs.\ (\ref{t12tky}), (\ref{varkapp})
for $T_{k_y}(E)$. Briefly speaking, the higher cumulant is considered,
the larger value of $k_FW$ is necessary to observe the convergence to the
sub-Sharvin limit; however, for $W/L=10$ and for the heavily-doped leads,
$k_FW\gtrsim{}5$ is sufficient.
The discussion of more realistic situations (finite doping in the leads,
other sample shapes) is presented later in this paper.

\subsection{The narrow-opening limit}
Using the conformal mapping technique \cite{Ryc09,Gui08}, it can be shown 
that for charge-neutral graphene ($k_F=0$), the pseudodiffusive values given
in Eqs.\ (\ref{sumpdiff}), (\ref{gfpdiff}), (\ref{r34pdiff}) are essentially
valid for an arbitrary sample shape, provided that the prefector $W/L$
(if present) is replaced by an appropriate geometry-dependent factor defined
by the conformal transformation.
In particular, when mapping the rectangle onto the Corbino
disk, one needs to substitute $W/L\rightarrow{}2\pi/\log(R_{\rm o}/R_{\rm i})$,
where $R_{\rm o}$ and $R_{\rm i}$ are the outer and inner disk radii
(respectively), see Fig.\ \ref{gfr34corb}(a).
An additional condition is that
the system must be in the multimode regime, i.e.\ 
$\log(R_{\rm o}/R_{\rm i})\ll{}1$ (or $R_{\rm i}\approx{}R_{\rm i}$) in the disk case.
Otherwise, if $R_{\rm i}\ll{}R_{\rm o}$, a~nonstandard tunneling behavior
is observed, with $G\propto{}R_{\rm i}/R_{\rm o}$ and $F\rightarrow{}1$
\cite{Ryc09}. (Since the transport is governed by a~single mode, we also have
$R_3\rightarrow{}1$ and $R_4\rightarrow{}1$ in such a~case.) 

However, in the Sharvin limit ($k_F{}R_{\rm i}\gg{}1$ in the disk case), 
a~different set of universal charge-transport characteristics is predicted
\cite{Ryc22}. Regardless of the exact size or shape of the outer sample-lead
interface, one can assume that the double-barrier formula,
Eq.\ (\ref{tt12phi}), is still applicable and that 
$T_2\approx{}1$ and $R_2=1-T_2\approx{}0$ due to the Klein tunneling.
Therefore, $T_{12}\approx{}T_1$, the role of a~phase shift $\phi$ is negligible,
and one can write --- for the wave leaving
the inner lead with the total angular momentum $\hbar{}j$ (with
$j=\pm{}1/2,\pm{}3/2,\dots$) --- the transmission probability as 
\begin{align}
  \left(T_j\right)_{R_{\rm i}\ll{}R_{\rm o}} &\approx{} T(u_j) =
    \begin{cases}
    {\displaystyle\frac{2\sqrt{1-u_j^2}}{1+\sqrt{1-u_j^2}}} \ \
      &\text{if }\ |u_j|\leq{}1, \\
    0\ \ &\text{if }\ |u_j|>1, 
  \end{cases} \nonumber\\
  \text{with }\ u_j &= \frac{j}{k_F{}R_{\rm i}}.
  \label{tujdef}
\end{align}
Subsequently, the summation for the $m$-th power of $T_j$ can be approximated
(notice that we have assumed $k_F{}R_{\rm i}\gg{}1$) by integration over
$-1\leqslant{}u\leqslant{}1$, leading to
\begin{equation}
  \sum_j(T_j)^m\approx{} k_F{}R_{\rm i}\int_{-1}^1{}du\left[T(u)\right]^m=
  2k_F{}R_{\rm i}\langle{}T^m\rangle_{u}, 
\end{equation}
where, also using the symmetry $T(-u)=T(u)$, we have defined 
\begin{widetext}
\begin{equation}
\label{avertmnarr}
  \langle{}T^m\rangle_{u} = \int_0^{1}du
  \left(\frac{2\sqrt{1-u^2}}{1+\sqrt{1-u^2}}\right)^m 
  =
  \frac{\sqrt{\pi}\,\Gamma(m+2)}{4\Gamma(m+\frac{5}{2})}\left[
    2m+3-2m\,{}_2F_1\left(\frac{1}{2},1;m+\frac{5}{2};-1\right) 
  \right]. 
\end{equation}
\end{widetext}
The first four values of $\langle{}T^m\rangle_{u}$ are listed
in Table~\ref{avtmtab}. 
Substituting the above into Eqs.\ (\ref{gland}), (\ref{fanodef}),
(\ref{r3def}), and (\ref{r4def}), we obtain
\begin{align}
  {G} &\approx (4-\pi)\,{G_{\rm Sharvin}},
  \nonumber \\ 
  &\ \text{with }\ {G_{\rm Sharvin}}=2g_0k_FR_{\rm i},
  \nonumber \\ 
  F &\approx \frac{9\pi-28}{3(4-\pi)} \simeq{} 0.1065,
    & (R_{\rm i}\ll{}R_{\rm o})
  \label{gfr34narrow} \\ 
  R_3 &\approx \frac{204-65\pi}{5(4-\pi)} \simeq{} -0.04742,
  \nonumber \\
  R_4 &\approx \frac{1575\pi-4948}{21(4-\pi)} \simeq{} 0.0004674.
  \nonumber 
\end{align}

For the Corbino disk, it is also possible to perform the analytical mode
matching for the heavily-doped leads, but arbitrary disk doping $k_F$
and radii ratio $R_{\rm o}/R_{\rm i}$ \cite{Ryc09,Ryc20}.
We skip the details of the derivation here and just give the transmission
probabilities
\begin{equation}
\label{tjhank}
  T_j=
  \frac{16}{\pi^2{}k^2{}R_{\rm i}{}R_{\rm o}}\,
  \frac{1}{\left[\mathfrak{D}_{j}^{(+)}\right]^2
    + \left[\mathfrak{D}_{j}^{(-)}\right]^2},
\end{equation}
where
\begin{align}
\label{ddnupm}
  \mathfrak{D}_{j}^{(\pm)} &= \mbox{Im}\left[ 
    H_{j-1/2}^{(1)}(k_FR_{\rm i})H_{j\mp{}1/2}^{(2)}(k_FR_{\rm o})\right. \nonumber\\
    & \ \ \ \ \ \ \ \ \ \ 
    \pm \left.H_{j+1/2}^{(1)}(k_FR_{\rm i})H_{j\pm{}1/2}^{(2)}(k_FR_{\rm o})
    \right], 
\end{align}
and $H_\nu^{(1)}(\rho)$ [$H_\nu^{(2)}(\rho)$] is the Hankel function of the
first [second] kind. 
In Fig.\ \ref{gfr34corb}, we provide the comparison between the
values of the conductance and the next three charge cumulants obtained
by substituting the exact $T_j$-s given above into Eqs.\ (\ref{gland}),
(\ref{fanodef}), (\ref{r3def}), (\ref{r4def}), and performing the numerical
summation over $j$ with the approximate values given in Eq.\
(\ref{gfr34narrow}). It is easy to see that the radii ratio of
$R_{\rm o}/R_{\rm i}=5$ is sufficient to reproduce our predictions for the
narrow-opening limit, with good accuracy, typically starting from
$k_F{}R_{\rm i}=50-100$. (This time the higher cumulant is considered,
the faster the convergence.)

\begin{table*}[t]
\caption{ \label{avtmtab}
  The first four cumulants for the transmission probabilities,
  $\langle{}T^m\rangle$,
  for different transport regimes in graphene and the corresponding
  values of the four charge-transfer characteristics, see Eqs.\
  (\ref{gland}), (\ref{fanodef}), (\ref{r3def}), (\ref{r4def}). 
}
\begin{tabular}{r|cccc}
 \multicolumn{5}{c}{ } \\
\hline\hline
  & \multicolumn{4}{c}{{\bf Transport regime} (or approximation)} \\
  Cumulant$\ \ $ & $\ ${\em Pseudodiffusive,}$\ $ & $\ ${\em Sub-Sharvin,}$\ $
    & $\langle{}\,X\,\rangle_{\rho_{\rm approx}(T)}$, & {\em Narrow-opening,} \\
   & $\ k_F\!=\!0,\ W\!\gg{}\!L^{a)}\ $ & $\ k_FW\gg{}1^{b)}\ $
    & $\ $see Eq.\ (\ref{ttmrhoapp})$^{c)}\ $
    & $\ k_FR_{\rm i}\!\gg{}\!1,\ R_{\rm i}\!\ll{}\!R_{\rm o}^{d)}\ $\\ \hline
  $\langle{}T\rangle\ \ $ & $1$ & $\pi/4$ & $\pi/4$ & $4-\pi$ \\
  $\langle{}T^2\rangle\ \ $ & ${2}/{3}$ & ${7\pi}/{32}$
    & ${2}/{3}$ & $40/3-4\pi$ \\
  $\langle{}T^3\rangle\ \ $ & ${8}/{15}$ & ${51\pi}/{256}$
    & $3\pi/16$ & $192/5-12\pi$ \\
  $\langle{}T^4\rangle\ \ $ & ${16}/{35}$ & ${759\pi}/{4096}$
    & $8/15$ & $32\left(332/105-\pi\right)$ \\
    & & & & \\
  ${G}/{G_{\rm Sharvin}}$ & $\infty^{e)}$ & $\pi/4$ & $\pi/4$ & $4-\pi$ \\
  $F\ \ $ & $1/3$ & $1/8$ & $1-8/3\pi\simeq{}0.1512$
    & $(9\pi-28)/3(4-\pi)$ \\
  $R_3\ \ $ & $1/15$ & $-1/32$ & $5/2-8/\pi\simeq{}-0.04648$
    & $(204-65\pi)/5(4-\pi)$ \\
  $R_4\ \ $ & $-1/105$ & $-5/512$ &  $\ 10-472/15\pi\simeq{}-0.01615\ $
    & $\ (1575\pi-4948)/21(4-\pi)\ $ \\
%  & & & & \\
\hline\hline
\multicolumn{5}{l}{
Expressions for $\langle{}T^m\rangle$ with arbirary $m\geqslant{}1$
are given by: 
$^{a)}$Eq.\ (\ref{avertmdiff}); 
$^{b)}$Eqs.\ (\ref{tmincoh}), (\ref{avertmincoh});  
$^{c)}$Eq.\ (\ref{ttmrhoapp}); 
$^{d)}$Eq.\ (\ref{avertmnarr}).
} \\
\multicolumn{5}{l}{
$^{e)}$For $k_F=0$, $G=g_0W/\pi{}L$, with $g_0=4e^2/h$, coincides
with $G_{\rm Sharvin}=0$.  
} \\
\end{tabular}
\end{table*}

\section{Distributions of transmission probabilities}
\label{distraei}

A~compact and intuitive representation of the charge-transfer cumulants
discussed in the previous Section is provided within the distribution
function of transmission probabilities $\rho(T)$.
This function takes the simplest form when the transmission probability
can be expressed as a~monotonic function of the parameter $\lambda$,
i.e., $T=T(\lambda)$.
In such a~case, the probability density is defined by
\begin{equation}
\label{rhottdef}
  \rho(T) = \rho(\lambda)\left|\frac{d\lambda}{dT}\right|, 
\end{equation}
where $\rho(\lambda)$ is the number of transmission channels per unit of
$\lambda$ [here constant and determined by the appropriate quantization
rule, see Eqs.\ (\ref{kynw}), (\ref{kynpiw}), or (\ref{tujdef})], and
${d\lambda}/{dT}$ is the derivative of the inverse function $\lambda(T)$. 
In a~generic situation, the right-hand side in Eq.\ (\ref{rhottdef}) needs
to be replaced by the sum over the monotonicity intervals
of $T=T(\lambda)$. 

In the pseudodiffusive limit, $k_F=0$ and $W\gg{}L$, the transmission
probability given by Eq.\ (\ref{tky0}) immediately implies
\begin{equation}
  \rho_{\rm diff}(T) = \frac{W}{2\pi{}L}\,\frac{1}{T\sqrt{1-T}} =
  \frac{G_{\rm diff}}{2\pi\sigma_0}\,\frac{1}{T\sqrt{1-T}},
\end{equation}
where we recall the pseudodiffisive conductance $G=G_{\rm diff}$ given
in Eq.\ (\ref{gfpdiff}). The distribution $\rho_{\rm diff}(T)$
is visualized in Fig.\ \ref{stripttky}(c). 

Analyzing the sub-Sharvin transport, we now change the order of presentation
by switching to the disk geometry to point out that in the narrow-opening
limit, i.e., for $k_FR_{\rm i}\gg{}1$ and $R_{\rm i}\ll{}R_{\rm o}$,
the transmission probability $T(u_j)$ given by Eq.\ (\ref{tujdef}) leads
to another closed-form expression for the distribution, namely
\begin{equation}
  \rho_{R_{\rm i}\ll{}R_{\rm o}}(T)= \frac{G_{\rm Sharvin}}{g_0}\,
  \frac{T}{(2-T)^2\sqrt{1-T}}, 
\end{equation}
with $G_{\rm Sharvin}=2g_0k_FR_{\rm i}$ for a~circular lead.

In the case of parallel interfaces at a~distance $L$, see Fig.\
\ref{stripttky}(a), the description of the sub-Sharvin transport becomes
more complex, since the transmission probability $T_{k_y}(E)$, see Eqs.\
(\ref{t12tky}) and (\ref{varkapp}), is no longer a~monotonic function of
$k_y$. The distribution $\rho(T)$ obtained numerically for
%$W/L=10$ and
$k_FL=25$ is presented in Fig.\ \ref{stripttky}(e), where the continuous
$k_y$ corresponds to the $W\gg{}L$ limit. 
It can be noticed 
that each of the seven transmission minima [see Fig.\ \ref{stripttky}(d)]
produces a~distinct (integrable) singularity of $\rho(T)$
located at $0<T_{\rm min}<1$.
A~closed-form, asymptotic expression for $\rho(T)$ in the
$k_F\rightarrow{}\infty$ limit is missing; instead, 
we propose the approximation directly following from
$\{T_{k_y}\}_{\rm incoh}$ given by Eq.\ (\ref{ttkyincoh}), i.e.,
\begin{equation}
\label{rhottapp}
  \rho_{\rm approx}(T) = \frac{G_{\rm Sharvin}}{g_0}\,
  \frac{T}{\sqrt{1-T^2}}. 
\end{equation}
Subsequent approximations for the cumulants can be evaluated as
\begin{equation}
\label{ttmrhoapp}
  \left\langle{}T^m\right\rangle_{\rho_{\rm approx}(T)} =
  \frac{g_0}{G_{\rm Sharvin}}\int_0^1{}dT\,T^m\rho_{\rm approx}(T),
  \ \ \ \ m\geqslant{}1. 
\end{equation}
The numerical values for $m=1,\dots,2$ are listed in Table~\ref{avtmtab},
together with the corresponding approximations for the charge-transfer
cumulants $F$, $R_3$, and $R_4$, which are also depicted in Figs.\
\ref{gfr34strip}(b--d) [thick horizontal lines].
We notice that these values typically match the incoherent ones, obtained by
substituting Eq.\ (\ref{tmincoh}) into Eq.\ (\ref{avertmincoh}),
within the accuracy that allows unambiguous identification of the transport
regime. 
A~surprising exception is the case of $R_4$, for which the
proximity of the pseudodiffusive ($-1/105$) and incoherent ($-5/512$) values
is merely a~coincidence. 
[By definition, the conductance $G\approx(\pi/4)\,G_{\rm Sharvin}=
(k_FW/\pi)\langle{}T\rangle_{\rho_{\rm approx}(T)}$.] 

The functional forms of $\rho(T)$ derived in this Section,
along with a~selection of others previously reported in the literature,
can be found in Table~\ref{gfrhotab}.

\begin{table*}[t]
\caption{ \label{gfrhotab}
  Basic quantum-transport regimes in selected nanoscopic systems characterized
  by the conductance ($G$), the Fano factor ($F$), and statistical
  distribution of transmission probabilities $\rho(T)$. Remaining symbols
  are the Fermi wavenumber $k_F$, the conductance quantum $g_0=2e^2/h$
  for the two-dimensional electron gas (2DEG) or $4e^2/h$ for graphene,
  and the number of open channels $N_{\rm open}$. 
}
\begin{tabular}{llcccc}
 & & & & & \\
\hline\hline
 {\bf Transport}$\ $
 & \multirow{2}{*}{System} & \multirow{2}{*}{$G$} & \multirow{2}{*}{$F$}
 & \multirow{2}{*}{$\rho(T)$} & \multirow{2}{*}{$\ $Refs.$\ $} \\
 {\bf regime} & & & & & \\ \hline
  & & & & & \\
 {\em Standard} & Sharvin contact in 2DEG,
   &  \multirow{2}{*}{$\ G_{\rm Sharvin}=g_0{k_FW}/{\pi}\ $}
   &  \multirow{2}{*}{$\ 0\ $}
   &  \multirow{2}{*}{$\ N_{\rm open}\,\delta(1-T)\ $}
   & \multirow{2}{*}{\cite{Wee88,Gla88}} \\
 {\em ballistic} & width $W$  &  &  & &  \\
   & & & & & \\
 {\em (Pseudo)-} & diffusive conductor
   & \multirow{2}{*}{$g_0\ll{}G\ll{}G_{\rm Sharvin}$}
   &  \multirow{2}{*}{$\ 1/3\ $}
   & \multirow{2}{*}{$\ \displaystyle\frac{G}{2g_0}\frac{1}{T\sqrt{1-T}}\ $}
   & \multirow{2}{*}{\cite{Bee92,Nag92}} \\
 {\em diffusive}  & & & & & \\
%% & & & & & \\
  &  Charge-neutral graphene sample
  & \multirow{2}{*}{$\displaystyle\sigma_0\frac{W}{L}$,
  $\ \displaystyle\sigma_0=\frac{4e^2}{\pi{}h}$}
  & \multirow{5}{*}{$1/3$}
  & \multirow{5}{*}{$\ \displaystyle\frac{G}{2\pi{}\sigma_0}\frac{1}{T\sqrt{1-T}}\ $}
  & \multirow{2}{*}{\cite{Two06,Dan08}} \\
  & (width $W$, length $L$) &  &  & & \\
    & & & & & \\
    &  Charge-neutral graphene disk
    &  \multirow{2}{*}{$\displaystyle
    \frac{2\pi{}\sigma_0}{\ln{}\left(R_{\rm o}/R_{\rm i}\right)}$}
    &  & 
    & \multirow{2}{*}{\cite{Ryc09}} \\
    & (inner radii $R_{\rm i}$, outer radii $R_{\rm o}$) &  &  &  & \\
      & & & & & \\
  {\em Sub-Sharvin}$\ $  & Doped graphene sample
  & \multirow{2}{*}{$\displaystyle\frac{\pi}{4}\,G_{\rm Sharvin}$}
  & \multirow{2}{*}{$1/8$}
  & \multirow{2}{*}{$\displaystyle\approx\frac{G_{\rm Sharvin}}{g_0}
    \frac{T}{\sqrt{1-T^2}}\ $}
  & \cite{Ryc21b},  \\
    & (width $W$, length $L$) &  &  & & this~work \\
      & & & & & \\
    &  Doped graphene disk, $B=0$
    & \multirow{2}{*}{$\ \displaystyle\frac{\pi}{4}^{a)}\!
      < \frac{G}{{G_{\rm Sharvin}}^{b)}} < 4-\pi^{c)}\ $}
    & \multirow{2}{*}{$\ 1/8^{a)}\!>F>0.1065^{c)}\ $}
  & \multirow{2}{*}{$\displaystyle
    \frac{({G_{\rm Sharvin}}/{g_0})\,T}{(2-T)^2\sqrt{1-T}}^{c)}\ $}
    & \cite{Ryc22}, \\
    & (inner radii $R_{\rm i}$, outer radii $R_{\rm o}$) &  &  &  & this~work \\
    & & & & & \\
  & Doped graphene disk, & \multirow{2}{*}{$G\rightarrow{}0$}
  & \multirow{2}{*}{$0.5497^{a)}<F<1^{c)}$} & \multirow{2}{*}{---}
  & \multirow{2}{*}{\cite{Ryc24}} \\
      & $B\rightarrow{}B_{c,2}-^{d)}$ & & & & \\
    & & & & & \\
  {\em Chaotic}  & Symmetric cavity & $0<G<{G_{\rm Sharvin}}^{e)}$  & $1/4$
  & $\displaystyle\frac{2G}{\pi{}g_0}\frac{1}{T^{1/2}\sqrt{1-T}}$
  & \cite{Bee97} \\
    & & & & & \\
\hline\hline
\multicolumn{6}{l}{
  $^{a)}$Reached for $R_{\rm i}/R_{\rm o}\rightarrow{}1$.
  $^{b)}$Defined as $G_{\rm Sharvin}=2g_0k_F{}R_{\rm i}$. 
  $^{c)}$Reached for $R_{\rm i}/R_{\rm o}\rightarrow{}0$.
} \\
\multicolumn{6}{l}{
  $^{d)}$Magnetic field corresponding to the vanishing conductance,
  $B_{c,2}=2\,(\hbar/e)k_F/(R_{\rm o}-R_{\rm i})$.
} \\
\multicolumn{6}{l}{  
  $^{e)}$Defined via the opening width $w$; i.e., $G_{\rm Sharvin}=g_0k_F{}w/\pi$. 
} \\
\end{tabular}
\end{table*}

\begin{table*}[t]
\caption{ \label{sysparams}
  Detailed parameters of the systems studied numerically (see also Fig.\
  \ref{allsetups}). For each case, the main spatial dimension is also given
  in physical units. 
}
\begin{tabular}{lccc}
 \multicolumn{4}{c}{ } \\
\hline\hline
  {\bf System}$\ $  &  $\ $Defining parameters$\ $
  & $\,\ $System (sample) length, ${L_{\rm tot}}\,(L')^{a)}\,\ $
  &  $\ $No.\ of sites$^{b)}\ $
  \\ \hline
  {\em Constriction with} & $W_{\infty}=210\sqrt{3}\,a$
  & $254\,a\simeq{}62.5\,$nm & $105,452$ \\
  {\em zigzag edges} & $\ W=60\sqrt{3}\,a,\,\ L=104\,a\ $
  & $(\,\equiv{}L)$ & $(24,960)$ \\
\multicolumn{4}{c}{ } \\
  {\em Half-Corbino disk} & $W_{\infty}=700\,a$
  & $120\sqrt{3}\,a\simeq{}51.1\,$nm & $336,000$ \\
    & $R_2=4R_1=200\,a$
    & $(\,\equiv{}R_2\!-\!R_1)$ & $(136,035)$ \\
\multicolumn{4}{c}{ } \\
  {\em Circular quantum dot}$\ $
  & $\ \ W_{\infty}\!=\!210\sqrt{3}\,a,\,\ R\!=\!105\sqrt{3}\,a\ $
  & $512\,a\simeq{}126\,$nm & $320,881$ \\
  & $w=60\sqrt{3}\,a$ &  $(362\,a)$ & $(240,389)$ \\
\multicolumn{4}{c}{ } \\
  {\em Circular quantum dot}$\ $
  &  $\ \ W_{\infty}\!=\!210\sqrt{3}\,a,\,\ R\!=\!105\sqrt{3}\,a\ $
  & $512\,a\simeq{}126\,$nm & $301,148$ \\
  {\em with a~circular hole}
  & $w=60\sqrt{3}\,a,\,\ r=30\sqrt{3}\,a$ & $(362\,a)$ & $(220,656)$ \\
\hline\hline
\multicolumn{4}{l}{
  $^{a)}{L_{\rm tot}}$---the distance between semi-infinite leads; 
  $L'$---the distance between interfaces (given in parenthesis). 
} \\
\multicolumn{4}{l}{
  $^{b)}$Total no.\ of sites between the leads.
  (No.\ of sites with $V(x)>-V_{\infty}/2$ is given in parenthesis.) 
} \\
\end{tabular}
\end{table*}

\section{Numerical results and discussion}
\label{restba}

In this Section, we compare the predictions from analytical
theory for the charge-transfer cumulants (see Sec.\ \ref{lanbutgra}) with
the results of computer simulations of electron transport in selected
nanostructures in graphene shown in Fig.\ \ref{allsetups} (for the
parameters, see Table~\ref{sysparams}).
Signatures of the sub-Sharvin transport regime originate from multiple
scattering between interfaces separating weakly and heavily doped regions;
therefore, it is important to compare the results for different
crystallographic orientations of such interfaces.
We note that for the system of Fig.\ \ref{allsetups}(a) the interfaces are
parallel to the armchair direction,
for the system of Fig.\ \ref{allsetups}(c) the semicircular interfaces probe
all crystallographic orientations, while for the systems of Figs.\
\ref{allsetups}(d) and \ref{allsetups}(e) the interfaces are parallel to
the zigzag direction.

Since discrete structures carved out of a~honeycomb lattice exhibit
Fabry-P\'{e}rot type oscillations in all studied transport properties as
a~function of Fermi energy, and (typically) the higher the cumulant, the
larger the oscillation magnitude, we limit the forthcoming discussion to
the Landauer-B\"{u}ttiker conductance ($G$) and the Fano factor ($F$). 
It is also worth noting that the ratio of the former to the Sharvin
conductance ($G/G_{\rm Sharvin}$) accompanied by $F$ provides sufficient
information to unambiguously identify one of the basic quantum transport
regimes, see Table~\ref{gfrhotab}, if applicable.

\subsection{Tight binding model}
This part of the analysis starts from the tight-binding model of graphene,
with Hamiltonian
\begin{equation}
\label{hamtba}
  H = \sum_{i,j,s}t_{ij}c_{i,s}^{\dagger}c_{j,s}+\sum_{i,s}V_in_{is}, 
\end{equation}
where the indices $i$, $j$ run over sites in the honeycomb lattice
of carbon atoms, and $s=\uparrow,\downarrow$ is the spin up/down orientation. 
The hopping-matrix elements are given by
\begin{equation}
  \label{tijcases}
  t_{ij} = 
  \begin{cases}
  -t_0 & \text{if } i,j\ \text{are nearest-neighbors}, \\
  0 & \text{otherwise}, \\
  \end{cases}
\end{equation}
with $t_0=2.7\,$eV.
For the systems shown in Figs.\ \ref{allsetups}(a), \ref{allsetups}(d),
and \ref{allsetups}(e), the electrostatic potential energy $V_j=V(x_j)$
varies only along the main axis. It equals $-V_{\rm infty}$, with
$V_{\rm infty}=t_0/2=1.35\,$eV, in the leads and raise to $V_j=0$ in the sample
area. The abrupt potential increase at the sample-lead interface is
smoothed over the length $L_s$, according to the function
\begin{equation}
\label{thelsdef}
  \Theta_{L_s}(x) =
  \begin{cases}
    0  & \text{if }\ x<-L_s/2, \\
    \frac{1}{2}+\frac{1}{2}\sin(\pi{}x/L_s)
      &  \text{if }\ |x|\leqslant{}L_s/2, \\
    1  & \text{if }\ x>L_s/2. \\
  \end{cases}
\end{equation}
The potential barrier, composed of two steps at $x=x_1$ and $x=x_2\equiv{}
x_1+L$,
namely
\begin{equation}
\label{vpotth}
  V(x) =  V_{\infty}\left[\,\Theta_{L_s}(x-x_1)-
  \Theta_{L_s}(x-x_2)\,\right] -V_{\infty},
\end{equation}
is rectangular for $L_s=0$ [solid line in Fig.\ \ref{allsetups}(b)],
whereas it has a~sinusoidal shape for $L_s=L$ [dashed line].
For the half-disk shown in Fig.\ \ref{allsetups}(c), we simply take
$V_j=V(r_j)$, where $r_j$ is the radius in polar coordinates, 
with the same function $V(r)$ as in Eq.\ (\ref{vpotth}).
The interface positions ($x_1,x_2$) coincide with the ends of the central
(narrowest) part with parallel edges [see Fig.\ \ref{allsetups}(a)], the
inner/outer disk radii [Fig.\ \ref{allsetups}(c)], or with the neckings
limiting the dot region [Fig.\ \ref{allsetups}(d,e)].  
The remaining symbols in Eq.\ (\ref{hamtba}) are a~creation (annihilation)
operator for electron with spin $s$ at lattice site $i$, $c_{i,s}^{\dagger}$
($c_{i,s}$) and the particle-number operator $n_{is}=c_{i,s}^{\dagger}c_{i,s}$.
(Since the Hamiltonian (\ref{hamtba}) can be represented as the sum of the
two commuting terms, one for $s=\uparrow$ and the other for $s=\downarrow$,
it is sufficient to calculate the transport characteristics for one spin
direction and to multiply the results by the degeneracy factor $2$.) 

In the following, we consider $0\leqslant{}L_s\leqslant{}L$ only for the
constriction shown in Fig.\ \ref{allsetups}(a); once the effect of the
smooth potential barrier is identified, the discussion of the remaining
systems concentrates on the case of $L_s=0$ (i.e., abrupt step).

\begin{figure*}[t]
%%\includegraphics[width=0.45\linewidth]{data0/GvsEF420x254zcstr}
%%\includegraphics[width=0.45\linewidth]{data0/FvsEF420x254zcstr}
%% ==> ...
\includegraphics[width=\linewidth]{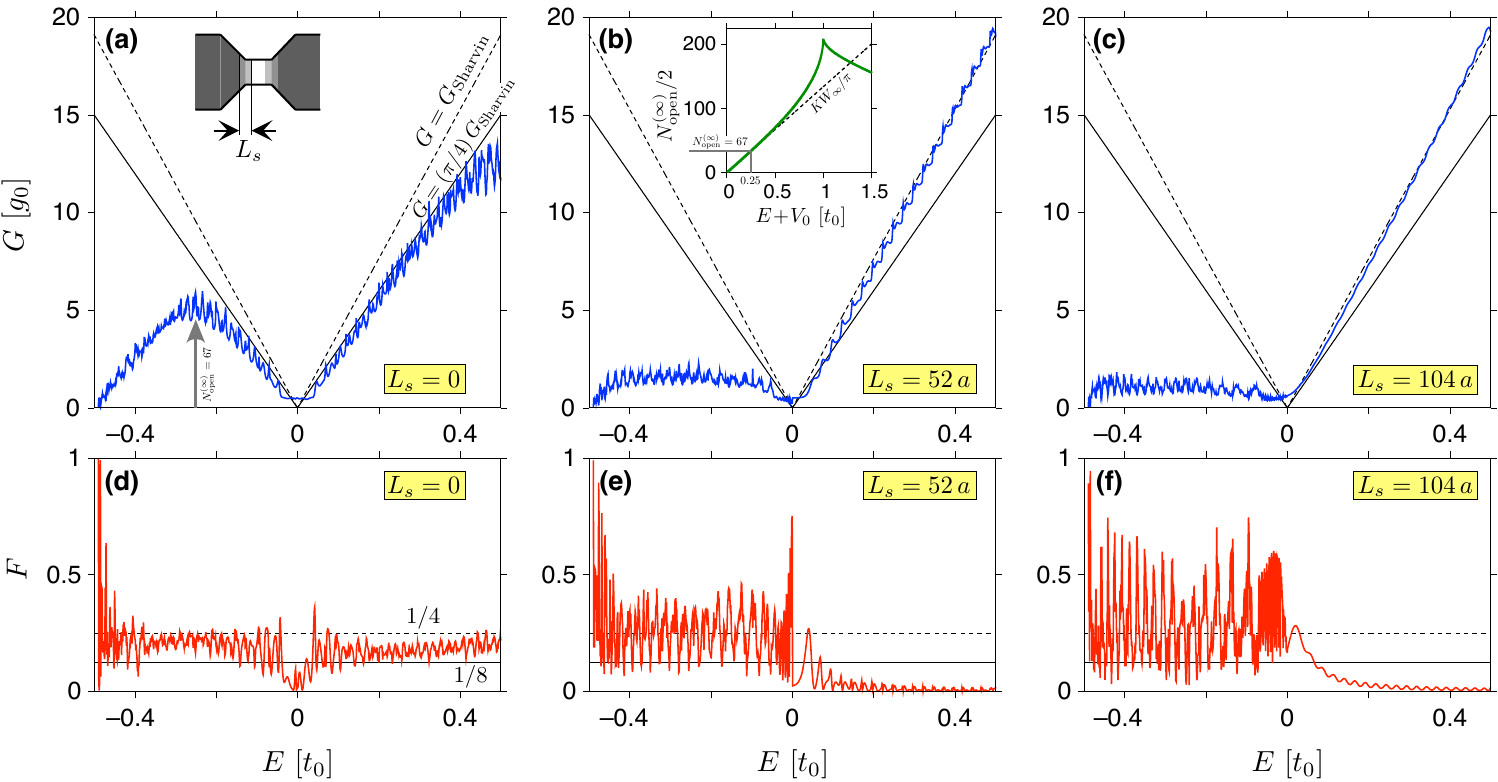}
\caption{ \label{gfzcstr}
  Conductance in the units of $g_0=4e^2/h$ (a--c) and the Fano factor
  (d--f) for the constriction with zigzag edges, see Fig.\ \ref{allsetups}(a),  
  displayed as functions of the Fermi energy defined with respect to the
  top ao the electrostatic potential barrier in the narrow region, see also
  Fig.\ \ref{allsetups}(b). 
  The subsequent panels correspond to abrupt ($L_s=0$), partly-smooth
  ($L_s=L/2=52\,a$), and fully-smooth ($L_s=L=104\,a$) potential steps.
  Numerical results following from the tight-binding calculations are
  depicted with thick lines. Thin solid lines mark the sub-Sharvin
  values given by Eqs.\ (\ref{gsubshar}), (\ref{fr34subshar}); dashed
  lines in (a--c) mark the Sharvin conductance given by Eq.\ (\ref{gshar})
  or, in (d--f), the shot-noise power characterizing symmetric cavity,
  $F=1/4$. 
  (The constriction width is $W=60\sqrt{3}\,a$; for the remaining simulation
  details, see Table~\ref{sysparams}.)
  Inset in (b) presents the number of propagating modes
  in the leads versus the energy $E+V_0$, with the step height
  $V_0=t_0/2=1.35\,$eV. 
}
\end{figure*}

\begin{figure*}[t]
\includegraphics[width=\linewidth]{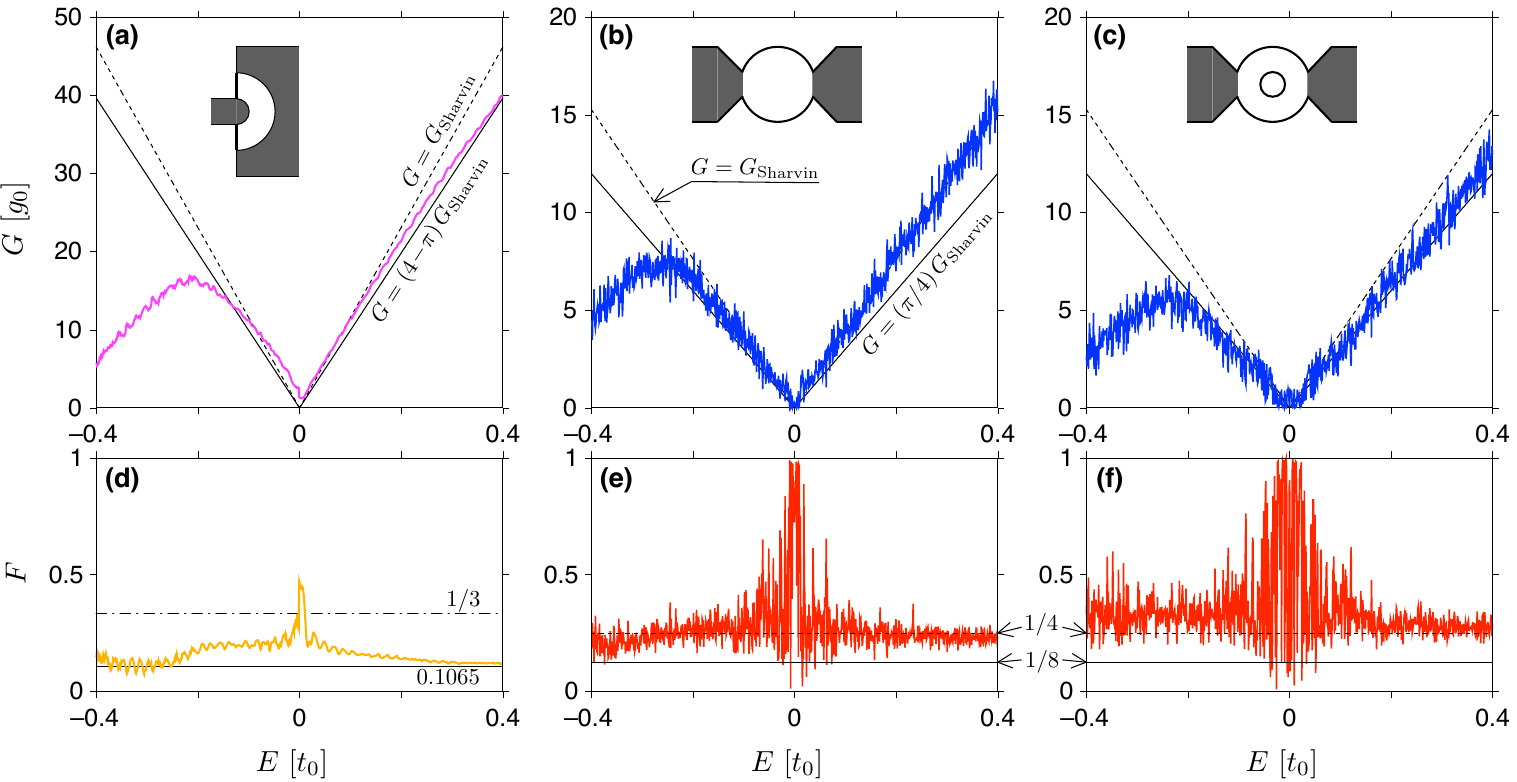}
\caption{ \label{gfcorgz}
  Same as in Fig.\ \ref{gfzcstr}, but for the half-Corbino disk (a,d)
  [see also Fig.\ \ref{allsetups}(c)], circular quantum dot (b,e)
  [see Fig.\ \ref{allsetups}(d)], and circular quantum dot with a~circular
  hole (c,f) [see Fig.\ \ref{allsetups}(e)].
  Thin solid lines in (a) and (d) show the results given in Eq.\
  (\ref{gfr34narrow}) for the narrow-opening limit;
  dash-dotted line in (d) marks the pseudodiffusive shot-noise power,
  $F=1/3$. Remaining lines are same as in Fig.\ \ref{gfzcstr}.
  (Other simulation details are given in Table~\ref{sysparams}.
}
\end{figure*}

Although electron-electron interaction is neglected in the tight-binding
model, Eq.\ (\ref{vpotth}) with $L_s>0$ can be regarded as a~simplified
description of carrier diffusion in a real device, leading to a~smooth
variation of the effective (self-consistent) potential in an interface
between regions of different doping \cite{Gio08,Cus17}. 
The possible role of electron-electron interactions in graphene has been
discussed by numerous authors
\cite{Mar97,Sor12,Sch13,Row14,Tan15,Rut23,Pot12},
leading to the conclusion that the correlation effects (not tractable within
the mean-field description) are typically negligible, and --- unless
extreme stains are applied \cite{Tan15,Rut23} ---
may manifest themselves only via partial magnetic order at the sample edges
\cite{Pot12}.

The nearest-neighbor hopping Hamiltonian (\ref{tijcases}) 
already grasps several features of more accurate models, such as the
trigonal warping of the dispersion relation and the presence of the
van Hove singularity in the density of states \cite{Ros20,Fab22}. 
For a~discussion of Hamiltonians with more distant hopping elements, 
see Ref.\ \cite{Vid22}.

\subsection{Constriction with zigzag edges}
As a~first example of the system, for which the analytical mode matching
technique presented in Sec.\ \ref{lanbutgra} cannot be directly applied, 
we consider the constriction with zigzag edges, earlier considered as
the valley \cite{Ryc07} or spin \cite{Wim08,Gru14} filter, depicted
in Fig.\ \ref{allsetups}(a).
The central section of this system is an almost perfect square, with the
length $L=104\,a\simeq{}25.58\,$nm and the width
$W=60\sqrt{3}\,a\simeq{}25.57\,$nm (see also Table~\ref{sysparams}),
attached to wedge-shaped electrodes that evolve into wide stripes with the 
width $W_{\rm infty}=210\sqrt{3}\,a\simeq{}89.5\,$nm.
Such a~geometry is chosen to mimic the typical experimental situation, in
which the nanostructure in graphene is contacted by much wider metallic leads
\cite{Bao10}. 
Also, the potential step height, $V_{\infty}=t_0/2\simeq{}1.35\,$eV is
not far from the results of some first-principles calculations for
graphene-metal structures \cite{Gio08,Cus17}.
Semi-infinite leads of a~constant width $W_{\infty}$ play a~role of the
waveguides shown in Fig.\ \ref{landauer:fig}; they can be divided into
the repeating supercells in order to find the propagating modes numerically,
by adapting the scheme developed by Ando for a~square lattice \cite{And91}
to the honeycomb lattice. For the potential profile given by Eqs.\
(\ref{thelsdef}) and (\ref{vpotth}), the number of propagating modes (per
one direction) is equal in the left and right leads, $N_L=N_R$
\cite{nmodfoo}.

Since the central section of the system is bounded by two parallel interfaces
separating weakly and heavily doped regions, one can expect that the key
findings for a~graphene strip in the sub-Sharvin regime, see Eqs.\
(\ref{gsubshar}) and (\ref{fr34subshar}) still apply,
at least for $L_s\ll{}L$.
However, the system width now varies with the position along the main axis,
so the scattering cannot be described independently for each normal mode,
as in Eq.\ (\ref{trphilphir}).
Instead, the mode mixing occurs, and --- if scattering from the left is
considered --- we can define the transmission matrix
${\boldsymbol t}=(t_{mn})$, with $m=1,\dots,N_R$ and $n=1,\dots,N_L$, and
the reflection matrix ${\boldsymbol r}=(r_{mn})$, with $m=1,\dots,N_L$,
$n=1,\dots,N_L$. 
The details of the calculations are presented 
in Appendix~\ref{nummodmat}; here we only mention that Eqs.\ (\ref{gland})
and (\ref{fanodef}) for measurable quantities remain valid, provided that
the transmission probabilities $T_n$ are defined as eigenvalues of the
matrix ${\boldsymbol t}{\boldsymbol t}^{\dagger}$.
Alternatively, one can express the Landauer-B\"{u}ttiker conductance and
the Fano factor in the basis-independent form,  referring to the traces of
the matrices ${\boldsymbol t}{\boldsymbol t}^{\dagger}$
and $\left({\boldsymbol t}{\boldsymbol t}^{\dagger}\right)^2$, namely
\begin{align}
  G &= \frac{2e^2}{h}\,\mbox{Tr}\left(
    {\boldsymbol t}{\boldsymbol t}^{\dagger}
  \right),
  \label{glandtr} \\
  F &= 1-\frac{\mbox{Tr}\left(
    {\boldsymbol t}{\boldsymbol t}^{\dagger}
  \right)^2}{\mbox{Tr}\left(
    {\boldsymbol t}{\boldsymbol t}^{\dagger}
  \right)}.
  \label{fanodeftr}
\end{align}
The factor $2$ in Eq.\ (\ref{glandtr}) denotes the spin degeneracy.
The valley degeneracy of the transmission eigenvalues is now only
approximate, since the dispersion relation following from the Hamiltonian
(\ref{hamtba}) is no longer perfectly conical, but shows the trigonal warping
\cite{Kat20}.
(For zigzag edges and electron doping, exact valley degeneracy occurs for
all but one mode; for armchair edges the degeneracy is approximate for all
modes \cite{Wak02,Wak10}.) 

The results of our computer simulations are depicted by the thick colored
lines in Fig.\ \ref{gfzcstr}.
They match the sub-Sharvin values (marked by black solid lines) for electron
doping ($E>0$) and the abrupt potential step ($L_s=0$).
For hole doping ($E<0$) and $L_s=0$ the conductance $G$ is still close
to $(\pi/4)\,G_{\rm Sharvin}$ as long as the number of propagating modes in the
leads is sufficiently large [see the inset in Fig.\ \ref{gfzcstr}(b)].
At the same time, the Fano factor is rather closer to the value of $F=1/4$, 
which characterizes the symmetric cavity.
In contrast, for smooth bariers ($L_s\gg{}a$) we have $G\approx{}G_{\rm Sharvin}$
for $E>0$ and $G\ll{}G_{\rm Sharvin}$ for $E<0$ (the conductance suppression
due to the presence of two p-n junctions), as can be expected for the 
standard (i.e., Sch\"{o}dinger) ballistic system.
At the same time, the Fano factor switches from $F\ll{}1$ (for $E>0$) to
$F\approx{}1/4$ (for $E<0$).
These findings are consistent with the results for smooth potential barriers
and a~strip with parallel edges, with mass confinement, presented
in Ref.\ \cite{Ryc21b}.

We see then that the constriction with zigzag edges
carved out of a~honeycomb lattice preserves all the key features of the
idealized Dirac system studied previously.

\subsection{Half-disk and circular quantum dots}
We now focus on the case of abrupt potential step ($L_s=0$) and consider
the geometries for which the possible role of the edges is reduced
(the half-Corbino disk) or amplified (circular quantum dot, without- or with
a~circular hole) compared to the constriction discussed above. 
It is worth to mention here that previous numerical studies on similar
nanostructures, either on the valley (or spin) filters
\cite{Ryc07,Wim08,Gru14} or the remaining systems \cite{Ryc09,Sch12,Ngu13}, 
have focused on the few-mode energy range, making it difficult or impossible
to distinguish between Sharvin and sub-Sharvin transport regimes.
(The same applies to the recent experimental study of graphene rings
\cite{Tan23}.) 
The conductance and the Fano factor determined from Eqs.\ (\ref{glandtr}) and
(\ref{fanodeftr}) after numerical calculation of the corresponding
transmission matrix (see also Appendix~\ref{nummodmat}) are presented in
Fig.\ \ref{gfcorgz}.  

In the half-disk case, see Figs.\ \ref{gfcorgz}(a) and \ref{gfcorgz}(d),
the conductance (for $E>0$) remains in the interval
$G_{\rm Sharvin}\gtrsim{}G\gtrsim{}(4-\pi)\,G_{\rm Sharvin}$
(notice that the radii ratio is $R_2/R_1=4$, and thus the relevant analytic
approximations are given in Eq.\ (\ref{gfr34narrow}) for the narrow-opening
limit),
with a~tendency to approach the narrow-opening value with increasing $E$. 
For $E<0$, the conductance behavior is less clear, but the values of $G$
are still close to both $G_{\rm Sharvin}$ and $(4-\pi)\,G_{\rm Sharvin}$.
In contrast, the Fano factor is close to the narrow-opening value of
$F\approx{}0.1065$ for both $E>0$ and $E<0$, except in the small vicinity
of the charge-neutrality point ($E=0$), where it is noticeably closer to
the pseudodiffusive value of $F=1/3$. 

For circular quantum dots Fabry-P\'{e}rot interference combined with 
scattering from irregular sample edges leads to much more pronounced 
oscillations of both $G$ and $F$, discussed as functions
of the Fermi energy, than in the case of a~half-Corbino disk.
In addition, the spectra presented in Figs.\ \ref{gfcorgz}(b,c) for $G$
and \ref{gfcorgz}(e,f) for $F$ suggest that the first charge-transfer
characteristic ($G$), discussed in isolation, may lead to the 
misidentification  of the Sharvin or sub-Sharvin transport regime.
Looking at the $F$ spectra, it is clear that the chaotic cavity
(with $F=1/4$) is the closest of the simple models that captures key features
of the circular quantum dot (both in the variant without- or with a~hole),
at least for higher electron or hole dopings. 
The conductance itself, related to the Sharvin value for $E>0$, appears to be
misleadingly close to $G_{\rm Sharvin}$ in the absence of a~hole, or to
$(\pi/4)\,G_{\rm Sharvin}$ in the presence of a~hole. 
(For $E<0$, the suppression of $G$ due to p-n junctions occurs in both cases.)

Therefore, complex nanostructures with irregular edges may accidentally
show some features of Sharvin (or sub-Sharvin) transport, but systematic
discussion of quantum transport unveils the chaotic nature of the system.

\section{Conclusions}
\label{conclu}

The main purpose of the work was to better understand the novel
{\em sub-Sharvin} transport regime in doped graphene, both by investigating
the analytical solutions for idealized systems and by comparing the results
for selected measurable quantities with computer simulations performed for
more realistic nanostructures.
For this goal,
we have developed the analytical technique that allows one to calculate
arbitrary charge transfer cumulant for doped graphene sample in two distinct
physical situations: 
(i) two long and parallel abrupt interfaces separate the sample and the leads;
(ii) a narrow circular interface governs transport through the much wider
sample toward an external lead.
In both cases, compact expressions are available for sufficiently high sample
doping (infinite doping is assumed for the leads), for which multiple
scattering between the interfaces can be taken into account, imposing
the random phase each time the electron passes the sample area.

We have also reviewed the most common quantum transport regimes described in
the literature, with their statistical distributions of transmission
probabilities. Evidence for a novel {\em sub-Sharvin} transport regime
in doped graphene is pointed out. 

Next, the results of analytical considerations for idealized systems are
compared with computer simulations of quantum transport for more realistic
systems carved out of a~honeycomb lattice.
The effects of finite doping
in the leads, smooth potential steps, trigonal warping, and irregular sample
edges are included in our simulations.
The results show that the main features of the analytical approach discussed
in the first part, which defines the sub-Sharvin transport regime in graphene
(with its variants for parallel interfaces and for the narrow-opening limit), 
are well reproduced in discrete systems on a~honeycomb lattice, provided
that the sample edges are straight and relatively short; i.e., with a~total
length comparable to or shorter than the total length of the sample-lead
interfaces.
In contrast, for systems with long and irregular edges, different
charge-transfer cumulants may suggest different quantum transport regimes,
making unambiguous classification difficult or impossible. 

Although our paper focuses on graphene, we expect that the main effects
will also occur in other two-dimensional crystals such as silicene,
germanene, or stanene \cite{Eza15,Har19}.
This prediction is based on the nature of the results presented, in particular
the fact that the occurrence of the sub-Sharvin transport regime is related
to the conical dispersion relation rather than to the transmission via
evanescent waves, which is responsible for the graphene-specific phenomena
that occur at the charge neutrality point. 

On the other hand, it seems unclear whether (or not) the sub-Sharvin transport
regime could appear in systems showing interplay between the conical and
quartic dispersion relations, such as bilayer graphene \cite{Cse07,Sus20b}, 
mono-bilayer junctions \cite{MDu21}, or hypothetical graphene-goldenene
junctions \cite{Kas24}.
These issues represent avenues for future theoretical and experimental
research.

\section*{Acknowledgements}
The work was mainly completed during a~sabbatical granted by the
Jagiellonian University in the summer semester of 2023/24. 
We gratefully acknowledge Polish high-performance computing infrastructure
PLGrid (HPC Center: ACK Cyfronet AGH) for providing computer facilities
and support within computational grant no.\ PLG/2024/017208.

%\noindent $\Rightarrow$
%{\sf Dalej --- DODATEK ze szczeg\'o\l{}owym wyprowadzeniem metody
%dla elektrod zigzag/armchair
%[Segregator: ``GRAPHENE\ II''; teczka: ``GRAPHENE: lattice vs continuous'']
%}

%%%%%%%%%%%%%%%%%%%%%%%%%%%%%%%%%%%%%%%%%%%%%%%%%%%%%%%%%%%%%%%%%%%%%%%%%%%%%%
\appendix

\begin{figure*}[t]
\includegraphics[width=\linewidth]{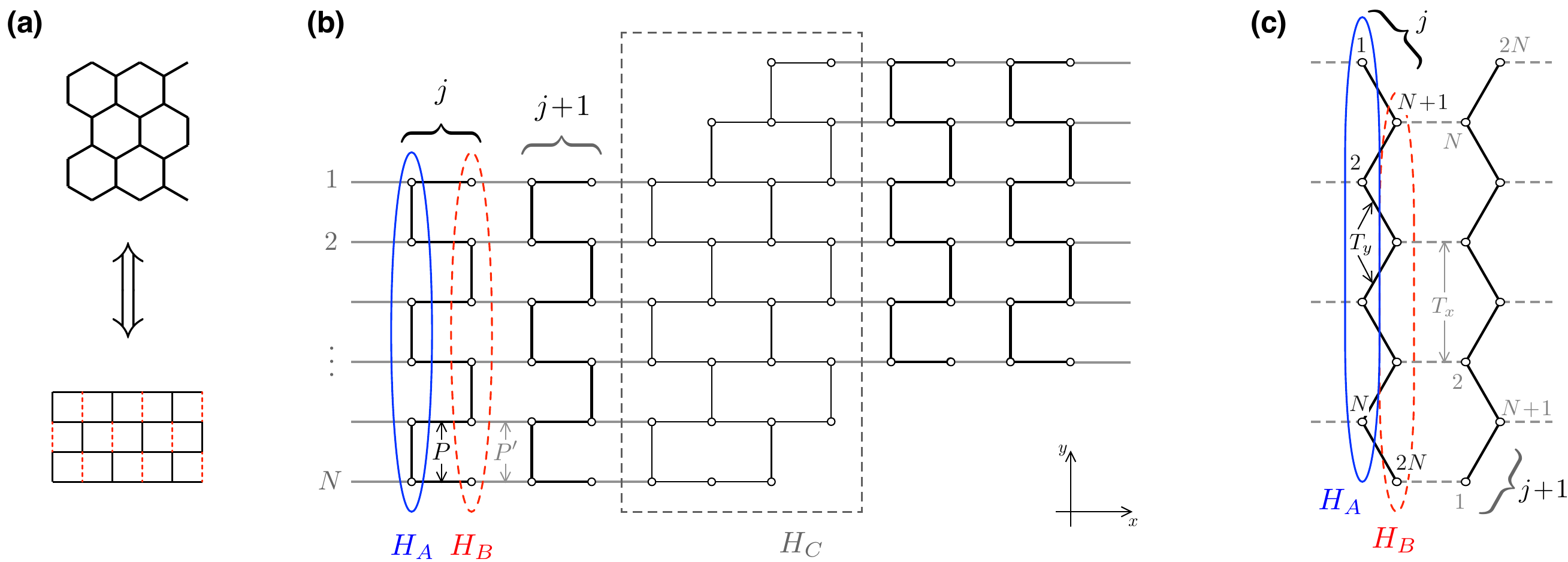}
\caption{ \label{squhon}
  (a) A~nanosystem of $24$ sites carved out of the honeycomb lattice
  and an equivalent section of a~square lattice with every second vertical
  bond removed (red dashed lines).
  (b) Schematic of an {\em open} system in computer simulation.
  Semi-infinite leads (with {\em zigzag} edges) are build from identical
  sections (the $j$-th and $j+1$
  sections are marked for the left lead), each consisting of two subsections
  $A$ and $B$, coupled only by horizontal bonds contributing to the matrix
  blocks $P$ and $P'$. Leads contact the
  central section (surrounded by dashed rectangle) of a~generic shape.
  (c) Analogous decomposition and the numbering scheme for the lead with
  {\em armchair edges}. 
  [See main text for details.] 
}
\end{figure*}

\section{Numerical mode-matching for the honeycomb lattice}
\label{nummodmat}

\subsection{The nanosystem}
In the following, we show how the computational method originally developed
by Ando for a square lattice \cite{And91} can be adapted to the honeycomb
lattice. 
The key point is to notice that --- when discussed on the level of
tight-binding Hamiltonian (\ref{hamtba}) --- a~honeycomb lattice with
nearest-neighbor hoppings only becomes equivalent to a~square lattice with
some bonds removed (see Fig.\ \ref{squhon}(a)).
The reasoning presented here can, with simple modifications, be applied to
a~honeycomb, square, or even triangular lattice (the last case
may be applicable when discussing the recently discovered two-dimensional
form of gold, {\em goldenene\/}, see Ref.\ \cite{Kas24}). 

The nanosystem, shown schematically in Fig.\ \ref{squhon}(b), represents
a~generic case that is tractable within the Landauer-B\"{u}ttiker formalism.
Two semi-infinite leads are built from repeatable sections, allowing one
to find the normal modes (propagating and evanescent) that define the basis
for the scattering matrix --- with moderate computational effort --- via
the secular equation (see next subsection);
the central part may have an arbitrary shape, and thus the mode matching
must be performed numerically on the lattice, which reduces to solving
a~sparse linear system of equations.

The details (and remaining assumptions) of both computational steps are
given below.

\subsection{Solutions in the leads}
Keeping in mind the analogy between the honeycomb lattice and the square
lattice mentioned above, we now consider an infinitely long wire with
a~width of $N$ sites, short sections of which are schematically shown in
Figs.\ \ref{squhon}(b) (for zigzag edges) and \ref{squhon}(c) (for armchair
edges).
Quantum-mechanical equation of motion can be written as 
\begin{equation}
\label{leadqueq}
  T^{\dagger}\varphi_{j-1}+H_0\varphi_j+T\varphi_{j+1}=E\varphi_j,
\end{equation}
where
\begin{equation}
  \varphi_j = \begin{pmatrix}
           \varphi_{A,j} \\
           \varphi_{B,j}
         \end{pmatrix}    
\end{equation}
is the $2N$-component wavefunction with the probability amplitudes
corresponding to the sites in the $A$ and $B$ blocks; i.e, 
$\varphi_{A(B),j}=\left[\varphi_{A(B),j}^{(1)},\dots,\varphi_{A(B),j}^{(N)}\right]^T$.
It is worth noting that the lead width in physical units, $W_\infty$, is
related to $N$ as follows 
\begin{equation}
  W_\infty = Na \times 
    \begin{cases}
    \frac{1}{2}\sqrt{3} & (\text{zigzag edges}), \\
    1 & (\text{armchair edges}). 
  \end{cases}
\end{equation}
Therefore, $W_\infty$ can be interpreted as a~circumference of a nanotube that
can be created by connecting the edge sites at each section by additional
(vertical) hopping. 

For {\em zigzag edges}, $H_0$ and $T$ are $2N\times{}2N$ matrices with the 
following block structure
\begin{equation}
  H_{0}^{({\rm zig})} = \begin{pmatrix}
           H_A^z & P \\
           P^{\dagger} & H_B^z
         \end{pmatrix},
         \ \ \ \ \ \ 
  T^{({\rm zig})} = \begin{pmatrix}
           0 & 0 \\
           P' & 0
         \end{pmatrix}.          
\end{equation}
The block matrices $H_A^z\neq{}H_B^z$ contain values of the electostatic
potential energy (i.e., $-V_0$ for all sites) and {\em vertical} hopping
elements depicted in Fig.\ \ref{squhon}; 
they are tridiagonal matrices with every second hopping removed, 
namely
\begin{align}
  \left(H_A^z\right)_{ll'} &=
    -V_0\delta_{l,l'}-t_0\left(\delta_{l,l'-1}+\delta_{l-1,l'}\right)
    [l\ \text{mod}\ 2], 
  \\
  \left(H_B^z\right)_{ll'} &=
    -V_0\delta_{l,l'}-t_0\left(\delta_{l,l'-1}+\delta_{l-1,l'}\right)
    [(l+1)\ \text{mod}\ 2], 
\end{align}
where $l,l'=1\dots{}N$, and $\delta_{ll'}$ is the Kronecker delta. 
The matrices $P$ and $P'$ contain horizontal hoppings; 
for zero magnetic field, we have
\begin{equation}
  \left(P\right)_{ll'} = \left(P'\right)_{ll'} = -t_0\delta_{l,l'}. 
\end{equation}
For a~uniform magnetic field, the Peierls substitution can be applied;
it reads
$P_{ll}\rightarrow{}P_{ll}\exp(2\pi{}il\Phi_a/\Phi_0)$ ($l=1\dots{}N$), with
$\Phi_a$ the flux per unit cell and $\Phi_0=h/e$ the flux quantum. 
(The generalization for some geometric strains, leading to position-dependent
hopping is also possible, provided that the invariance upon
$j\rightarrow{}j+1$, see Eq.\ (\ref{leadqueq}), is preserved.) 

For {\em armchair edges}, the site-numbering scheme presented in Fig.\
\ref{squhon}(c) allows us to keep the block structure of Eq.\
(\ref{leadqueq}), with
\begin{equation}
\label{armblocks}
  H_{0}^{({\rm arm})} = \begin{pmatrix}
           H_A^a & T_y \\
           T_y^{\dagger} & H_B^a
         \end{pmatrix},
         \ \ \ \ \ \ 
  T^{({\rm arm})} = \begin{pmatrix}
           0 & 0 \\
           T_x & 0
         \end{pmatrix}.          
\end{equation}
Now, the matrices $H_A^{a}=H_B^{a}$ contain only diagonal elements,
\begin{equation}
  \left(H_A^a\right)_{ll'} = \left(H_B^a\right)_{ll'} = -V_0\delta_{l,l'}. 
\end{equation}
The remaining blocks, $T_x$ and $T_y$, are given by 
\begin{align}
  \left(T_x\right)_{ll'} &= -t_0\delta_{l,N-l'+1}
  \\
  \left(T_y\right)_{ll'} &= -t_0\left(\delta_{l,l'} + \delta_{l-1,l'} \right),  
\end{align}
where $l,l'=1\dots{}N$ again.

Using the familiar Bloch ansatz,
\begin{equation}
  \varphi_j=\lambda^j\varphi_0, 
\end{equation}
we arrive to the secular equation, for zigzag edges, 
\begin{equation}
\label{secueq}
  \begin{pmatrix}
    H_A^z\!-\!E\mathbb{I} & P \\
    P' & 0 
  \end{pmatrix}^{-1}
  \begin{pmatrix}
    0 & -{P'}^{\dagger} \\
    -P^{\dagger} & E\mathbb{I}\!-\!H_B^z 
  \end{pmatrix}
  \varphi_0=\lambda\varphi_0,  
\end{equation}
where $\mathbb{I}$ is the $N\times{}N$ identity matrix and 
the eigenvalues $\lambda$ are complex with $|\lambda|=1$
(for {\em propagating modes}) or
real with $|\lambda|\neq{}1$ (for {\em evanescent} modes). 
For armchair edges, Eq.\ \ref{secueq} holds with the following
substitutions 
\begin{equation}
\label{secueqarm}
  H_A^z \rightarrow{} H_A^a, \ \ \ H_B^z \rightarrow{} H_B^a, \ \ \
  P \rightarrow T_y, \ \ \ P'\rightarrow{} T_x. 
\end{equation}
(If the generalization involving the dependence of the matrix elements 
on the position across the lead, such as the Peierls substitution mentioned 
above, is desired, one must double the number blocks in Eq.\
(\ref{armblocks}) by introducing 
$\varphi_j=(\varphi_{A1,j},\varphi_{B1,j},\varphi_{A2,j},\varphi_{B2,j})^T$; 
however, such a~case is beyond the scope of the present paper).

\subsection{Basis for the scattering matrix}
The secular equation derived above, given explicitly by Eq.\ (\ref{secueq})
for zigzag edges (z.e.), with the necessary substitutions for armchair edges
(a.e.) in Eq.\ (\ref{secueqarm}), can be diagonalized numerically using 
standard software packages in order to find the full set of eigenvalues
($\lambda$) with corresponding right-eigenvectors $\varphi_0^{(\lambda)}$.
We chose the double precision LAPACK
routines {\tt dgemm} and {\tt dgeev}, see Ref.\ \cite{lapack99}. 

%$\Rightarrow\,$ 
%{\sf Uwaga o~tym, jak ustalany jest kierunek pr\k{a}du i~normalizacja}
%[PATRZ:
% ---plik {\tt graver.f90}, procedury {\tt find\_vka}, {\tt iter\_jpa};  
% ---plik {\tt grazkl.f90}, ... ]

To correctly define the scattering matrix, the propagating modes must be
normalized so that they carry equal currents along the main axis of each
lead (i.e., the $x$-axis direction in Fig.\ \ref{squhon}). 

In general, the total current incoming at the $i$-th lattice site of
the system described by the tight-binding Hamiltonian, see Eqs.\
(\ref{hamtba}) and (\ref{tijcases}) in the main text, is given by the time
derivative of the charge, 
\begin{equation}
  \dot{n}_i = i[H,n_i]=  i\sum_{j(s)}
  \left(t_{ij}c_i^{\dagger}c_j-t_{ij}^{\star}c_j^{\dagger}c_i\right), 
\end{equation}
where $j(i)$ runs over the nearest neighbors of $i$, complex hopping 
(i.e., $t_{ij}\neq{}t_{ij}^\star$) is allowed, and the spin is omitted 
for clarity. In turn, the current flowing from site $i$ to site $j$ is 
described by the quantum-mechanical operator 
\begin{equation} 
  J_{ij} = i\left(t_{ij}c_i^{\dagger}c_j-t_{ij}^{\star}c_j^{\dagger}c_i\right). 
\end{equation} 
Taking into account all the currents incoming and outgoing from
a~repeatable section ($j$) of the lead (see Fig.\ \ref{squhon}) via individual
bonds, projected onto the $x$-direction, bring us to the total $x$-current
operator $J_x$, which --- when acting on the right-eigenvector
$\varphi_0^{(\lambda)}$, see Eqs.\ (\ref{secueq}) and (\ref{secueqarm}), 
corresponding to the eigenvalue $\lambda$ --- is equivalent to
\begin{equation}
  J_x^{(\lambda)} = i 
%    \begin{cases}
    \begin{pmatrix}
      0 & -\lambda^{-1}{P'}^{\dagger} \\ \lambda{}P' & 0
    \end{pmatrix}
      \ \ \ \  (\text{z.e.}), 
\end{equation}
or
\begin{equation}
  J_x^{(\lambda)} = i 
    \begin{pmatrix}
      0 & -\lambda^{-1}T_x^{\dagger}  \\ \lambda{}T_x & 0
    \end{pmatrix}
      \ \ \ \  (\text{a.e.}). 
%  \end{cases} 
\end{equation}
%where the block structure is adjusted to the site-numbering scheme.

Subsequently, the normalized eigenvector can be written as
\begin{equation}
  {\boldsymbol v}_\lambda^{\rm pro} =
  \frac{1}{\sqrt{\left|{\varphi_0^{(\lambda)}}^{\dagger}J_x^{(\lambda)}\varphi_0^{(\lambda)}\right|}}
  \,\varphi_0^{(\lambda)}\ \ \ \
  (\text{for }\ |\lambda|=1),  
\end{equation}
and the sign 
\begin{equation}
%  \mbox{sgn}\left\langle{}J_x\right\rangle_\lambda
  s_\lambda 
  \equiv
  \left({\boldsymbol v}_\lambda^{\rm pro}\right)^{\dagger}J_x^{(\lambda)}
  {\boldsymbol v}_\lambda^{\rm pro} 
   = \pm{}1 
\end{equation}
indentifies the direction of propagation. 
For evanescent modes, the normalization is irrelevant, and we simply set 
\begin{equation}
  {\boldsymbol v}_\lambda^{\rm eva} = \varphi_0^{(\lambda)}\ \ \ \
  (\text{for }\ |\lambda|\neq{}1). 
\end{equation}
In analogy to propagating modes, $|\lambda|>1$ now identifies the 
evanescent mode decaying to the left, while $|\lambda|<1$ identifies 
the evanescent mode decaying to the right. 
The real-space components of the normalized eigenvector for the $j$-th
section of the lead can be written as column vectors of the dimension $2N$,
namely 
\begin{equation}
  {\boldsymbol v}_{\lambda,j}^{{\rm pro}}=
  \lambda^j{\boldsymbol v}_{\lambda}^{{\rm pro}},
  \ \ \ \ \ \ 
  {\boldsymbol v}_{\lambda,j}^{\rm eva}= \lambda^j{\boldsymbol v}_{\lambda}^{\rm eva}.  
\end{equation}

Now it is worth generalizing the discussion a~bit by allowing that the left
and right leads are not necessarily identical, 
i.e., may have different numbers of sites across, $N^{(L)}$ and $N^{(R)}$,
or different electostatic potential energy levels, $-V_0^{L}$ and $-V_0^{R}$.
In such a~case, the secular equation, Eq.\ \ref{secueq}, appears in the
two versions, with the block matrices replaced by
\begin{align}
  H_A&\rightarrow{}H_A^{\alpha},\ \  H_B\rightarrow{}H_B^{\alpha},\ \ 
  P\rightarrow{}P_\alpha, \ \ {P'}\rightarrow{}{P_\alpha'},
  \nonumber\\
  &\text{for }\ \alpha=L,R,  \label{secueqlr}
\end{align}
leading to ({\em a~priori\/}) different numbers of propagating modes per
single direction, $N_{\rm pro}^{(L)}$ and $N_{\rm pro}^{(R)}$.
The number of left- (or right-) decaying evanescent modes is then given by
\begin{equation}
  N_{\rm eva}^{(\alpha)} = N^{(\alpha)}-N_{\rm pro}^{(\alpha)}, \ \ \ \
  \text{for }\ \alpha=L,R. 
\end{equation}
Therefore, the number of elements (i.e., normalized eigenvectors)
for the following four sets of (left-, right-) propagating and (left-, right-)
decaying modes in a~single electrode are as follows 
\begin{align}
  \#\left\{{\boldsymbol v}_{\lambda,j}^{\alpha,{\rm pro}}, s_\lambda >0\right\} &=
  \#\left\{{\boldsymbol v}_{\lambda,j}^{\alpha,{\rm pro}}, s_\lambda<0\right\} =
  N_{\rm pro}^{(\alpha)}, \\
  \#\left\{{\boldsymbol v}_{\lambda,j}^{\alpha,{\rm eva}}, |\lambda|>1\right\} &=
  \#\left\{{\boldsymbol v}_{\lambda,j}^{\alpha,{\rm eva}}, |\lambda|<1\right\} =
  N^{(\alpha)}\!-\!N_{\rm pro}^{(\alpha)}, \\
  & \text{for }\ \alpha=L,R. \nonumber
\end{align}

For further considerations, we define the matrices $W_j^{(L)}$ and $W_j^{(R)}$,
in which the selected eigenvectors are stored column by column, i.e.,
\begin{widetext}
\begin{equation}
  W_j^{(L)} = \begin{pmatrix} W_{A,j}^{(L)} \\ W_{B,j}^{(L)} \end{pmatrix} = 
  \left(
    \left\{{\boldsymbol v}_{\lambda,j}^{L,{\rm pro}}, s_\lambda >0\right\};
    \left\{{\boldsymbol v}_{\lambda,j}^{L,{\rm pro}}, s_\lambda<0\right\};
    \left\{{\boldsymbol v}_{\lambda,j}^{L,{\rm eva}}, |\lambda|>1\right\}
  \right) \equiv
  \begin{pmatrix}
    V_{A,j}^{(L)} & U_{A,j}^{(L)} \\
    V_{B,j}^{(L)} & U_{A,j}^{(L)}
  \end{pmatrix}, 
\end{equation}
and 
\begin{equation}
  W_j^{(R)} = \begin{pmatrix} W_{A,j}^{(R)} \\ W_{B,j}^{(R)} \end{pmatrix} = 
  \left(
    \left\{{\boldsymbol v}_{\lambda,j}^{R,{\rm eva}}, |\lambda|<1\right\}; 
    \left\{{\boldsymbol v}_{\lambda,j}^{R,{\rm pro}}, s_\lambda >0\right\};
    \left\{{\boldsymbol v}_{\lambda,j}^{R,{\rm pro}}, s_\lambda<0\right\}
  \right)\equiv
  \begin{pmatrix}
    U_{A,j}^{(R)} & V_{A,j}^{(R)} \\
    U_{B,j}^{(R)} & V_{A,j}^{(R)}
  \end{pmatrix}. 
\end{equation}
\end{widetext}
In the above, the blocks $W_{A,j}^{(\alpha)}$ and $W_{B,j}^{(\alpha)}$ store
the matrix rows $1\dots{}N^{(\alpha)}$ and $N^{(\alpha)}+1\dots{}2N^{(\alpha)}$
(respectively), for $\alpha=L,R$.
These blocks correspond to the first $N^{(\alpha)}$
sites ($W_{A,j}^{(\alpha)}$), and to the next $N^{(\alpha)}$ sites
($W_{B,j}^{(\alpha)}$) in the $j$-th section of a~single lead. 

Now, the real-space components of an arbitrary wavefunction in the $j$-th
section of the (left,right) lead can be represented as 
\begin{align}
  \Psi_j^{(L)} = W_j^{(L)}\left(
    \left\{a_{\lambda,+}^{L,{\rm pro}}\right\};
    \left\{b_{\lambda,-}^{L,{\rm pro}}\right\};
    \left\{b_{\lambda,-}^{L,{\rm eva}}\right\}
  \right)^T, \\
  \Psi_j^{(R)} = W_j^{(R)}\left(
    \left\{b_{\lambda,+}^{R,{\rm eva}}\right\};
    \left\{b_{\lambda,+}^{R,{\rm pro}}\right\};
    \left\{a_{\lambda,-}^{R,{\rm pro}}\right\}
  \right)^T, 
\end{align}
where each set of complex coefficients,
$\left\{a_{\lambda,+}^{L,{\rm pro}}\right\}$, etc.,
corresponds to the matching set of eigenvectors in the matrix $W_j^{(L)}$
or $W_j^{(R)}$. 
Parts of the matrices $W_j^{(\alpha)}$, $\alpha=L,R$, containing the columns
corresponding to the coefficients $b_{\lambda,\mp}^{\alpha,{\rm pro}}$,
$b_{\lambda,\mp}^{\alpha,{\rm eva}}$, read 
\begin{align}
  \begin{pmatrix}
    U_{A,j}^{(L)} \\ U_{B,j}^{(L)}
  \end{pmatrix}
  &= \left(
    \left\{{\boldsymbol v}_{\lambda,j}^{L,{\rm pro}}, s_\lambda<0\right\};
    \left\{{\boldsymbol v}_{\lambda,j}^{L,{\rm eva}}, |\lambda|>1\right\}
  \right), \\
  \begin{pmatrix}
    U_{A,j}^{(R)} \\ U_{B,j}^{(R)} 
  \end{pmatrix}
  &= \left(
    \left\{{\boldsymbol v}_{\lambda,j}^{R,{\rm eva}}, |\lambda|<1\right\}; 
    \left\{{\boldsymbol v}_{\lambda,j}^{R,{\rm pro}}, s_\lambda >0\right\}
  \right),  
\end{align}
where each of the blocks $U_{A,j}^{(\alpha)}$, $U_{B,j}^{(\alpha)}$
is a~square matrix of dimension $N^{(\alpha)}\times{}N^{(\alpha)}$.
The remaining columns, corresponding to the coefficients
$a_{\lambda,\pm}^{\alpha,{\rm pro}}$, define the matrices
\begin{align}
  \begin{pmatrix}
    {V}_{A,j}^{(L)} \\ {V}_{B,j}^{(L)}
  \end{pmatrix}
  &= \left(
    \left\{{\boldsymbol v}_{\lambda,j}^{L,{\rm pro}}, s_\lambda>0\right\}
  \right), \\
  \begin{pmatrix}
    {V}_{A,j}^{(R)} \\ {V}_{B,j}^{(R)} 
  \end{pmatrix}
  &= \left(
    \left\{{\boldsymbol v}_{\lambda,j}^{R,{\rm pro}}, s_\lambda <0\right\}
  \right),    
\end{align}
where the dimensions of the blocks $V_{A,j}^{(\alpha)}$, $V_{B,j}^{(\alpha)}$ are
$N^{(\alpha)}\times{}N_{\rm pro}^{(\alpha)}$, for $\alpha=L,R$.

\subsection{The scattering problem}
In the last part of this Appendix, we will use the notation relevant
for the leads with zigzag edges (see Fig.\ \ref{squhon}).
Nevertheless, if a~version of the approach for the leads with armhair edges
is desired, it can be easily generated by performing the substitutions
given by Eq.\ \ref{secueqarm}.
(We emphasize here that the central section of the system,
described by the tight-binding Hamiltonian $H_C$, may have an
arbitrary shape, so that the irregular edges in the central section are also
tractable within the presented approach.)

Using the definitions introduced in the previous subsections, we can now
write down the quantum-mechanical equation of motion for the entire
nanosystem as follows
\begin{widetext}
\begin{equation}
\label{bigmatrixrec}
  \begin{pmatrix}
    {P_L}^{\dagger}{W}_{A,-1}^{(L)}+(H_B^{L}\!-\!E){W}_{B,-1}^{(L)} &
    \ \ \overrightarrow{P_L'}\ \  & & &   \\
    {\overrightarrow{P_L'}}^{\dagger}{W}_{B,-1}^{(L)} & & & & \\
    & & H_C-E & & \\
     & & & & \overleftarrow{P_R'}{W}_{A,1}^{(R)} \\
     & & & \ \ {\overleftarrow{P_R'}}^{\dagger}\ \  &
     (H_A^{R}\!-\!E){W}_{A,1}^{(R)} + P_R{W}_{B,1}^{(R)} 
  \end{pmatrix}
  \begin{pmatrix}
    \left\{a_{\lambda,+}^{L,{\rm pro}}\right\} \\
    \left\{b_{\lambda,-}^{L,{\rm pro}}\right\} \\
    \left\{b_{\lambda,-}^{L,{\rm eva}}\right\} \\
  \Psi_C \\
    \left\{b_{\lambda,+}^{R,{\rm eva}}\right\} \\
    \left\{b_{\lambda,+}^{R,{\rm pro}}\right\} \\
    \left\{a_{\lambda,-}^{R,{\rm pro}}\right\} 
  \end{pmatrix}
  = 0, 
\end{equation}
\end{widetext}
where we set $j=-1$ for the terminal section of the left lead, 
and $j=1$ for the terminal section of the right lead. 
The unit matrix is omitted in expressions $(H_B^{L}-E\mathbb{I})$ and
$(H_A^{R}-E\mathbb{I})$; 
the coefficients in each set, $\left\{a_{\lambda,+}^{L,{\rm pro}}\right\}$, etc.,
are now listed as columns; 
$\Psi_C$ is a~column vector that stores the wavefunction
amplitudes for the central section.

%... {\sf JESZCZE trzeba wprowadzi\'c} $P_L$, $P_R$ {\sf bo elektrody s\k{a}
%r\'o\.zne}$\,$({\bf !!!}) ...
%{\sf Podobnie z} $H_B$, $H_B$ ...

In a~case when the width of the central section, $N_C$, exceeds the width of
(at least) one of the leads, i.e., $N_C>N^{(L)}$ or $N_C>N^{(R)}$ (in addition,
the leads may be vertically displaced, as in the nanosystem shown in Fig.\
\ref{squhon}), the blocks occurring near the upper-left or lower-right
corner of the main matrix in Eq.\ (\ref{bigmatrixrec}) that connect the
central section with the leads, namely,
$\overrightarrow{P_L'}$ and $\overleftarrow{P_R'}$ with their hermitian
conjugates ${\overrightarrow{P_L'}}^{\dagger}$, ${\overleftarrow{P_R'}}^{\dagger}$, 
are constructed in such a~way that the original matrix $P_\alpha'$
is horizontally or vertically expanded and filled with zeros, namely
\begin{equation}
  \overrightarrow{P_L'} = \begin{pmatrix} 0 & P_L'& 0 \end{pmatrix},
  \ \ \ \ \ \ 
  \overleftarrow{P_R'} = \begin{pmatrix} 0\\ P_R'\\ 0 \end{pmatrix},   
\end{equation}
where the elements in the original block match the existing
bonds connecting the lattice sites.
(For $N_C=N^{(L)}=N^{(R)}$, we simply have $\overrightarrow{P_L'}=P_L'$ and
$\overleftarrow{P_R'}=P_R'$.)

In addition, Eq.\ (\ref{secueq}) with substitutions given by Eq.\
(\ref{secueqlr}) for the two leads, guarantees that
\begin{align}
  {P_L}^{\dagger}{W}_{A,-1}^{(L)}+(H_B^L\!-\!E){W}_{B,-1}^{(L)} &= -P_L'{W}_{A,0}^{(L)}, 
  \label{simple-ul}
  \\
  (H_A^R\!-\!E){W}_{A,1}^{(R)} + P_R{W}_{B,1}^{(R)} &= -{P_R'}^{\dagger}{W}_{B,0}^{(R)},
  \label{simple-br}
\end{align}
allowing some further simplification of Eq.\ (\ref{bigmatrixrec}).

We now introduce the
$(N_{\rm pro}^{(L)}+N_{\rm pro}^{(R)})\times{}(N_{\rm pro}^{(L)}+N_{\rm pro}^{(R)})$ 
scattering matrix, 
\begin{equation}
\label{smatdef}
  S = \begin{pmatrix}
    r & t' \\
    t & r' 
  \end{pmatrix}, 
\end{equation}
defined in such a way that  
\begin{equation}
  \begin{pmatrix} \left\{b_{\lambda,-}^{L,{\rm pro}}\right\} \\
  \left\{b_{\lambda,+}^{R,{\rm pro}}\right\} \end{pmatrix}
   = S \begin{pmatrix} \left\{a_{\lambda,+}^{L,{\rm pro}}\right\} \\
  \left\{a_{\lambda,-}^{R,{\rm pro}}\right\} \end{pmatrix}. 
\end{equation}
If we assume that the coefficients $\left\{a_{\lambda,+}^{L,{\rm pro}}\right\}$
and $\left\{a_{\lambda,-}^{R,{\rm pro}}\right\}$ are independent amplitudes
of the incoming waves in the left and right leads, respectively,
we can transform the linear system in Eq.\ (\ref{bigmatrixrec}) into a~series
of Kramers systems, one for each incoming mode $l$; namely, 
\begin{widetext}
\begin{equation}
\label{kramsys}
  \begin{pmatrix}
%%    {P}^{\dagger}{V}_{A,-1}^{(L)}+(H_B\!-\!E){V}_{B,-1}^{(L)} &
    -P_L'{U}_{A,0}^{(L)} & 
    \ \ \overrightarrow{P_L'}\ \  & & &   \\
    {\overrightarrow{P_L'}}^{\dagger}{U}_{B,-1}^{(L)} & & & & \\
    & & H_C-E & & \\
     & & & & {\overleftarrow{P_R'}}{U}_{A,1}^{(R)} \\
     & & & \ \ {\overleftarrow{P_R'}}^{\dagger}\ \  &
%%     (H_A\!-\!E){V}_{A,1}^{(R)} + P{V}_{B,1}^{(R)}
     -{P_R'}^{\dagger}{U}_{B,0}^{(R)}
  \end{pmatrix}
  \begin{pmatrix}
    \ r & t' \\
    \multicolumn{2}{c}{\big\{\Psi_C^{(l)}\big\}}  \\
    \ t & r' 
  \end{pmatrix}
  =
  \begin{pmatrix}
  P_L'{V}_{A,0}^{(L)} & 0 \\
  -{\overrightarrow{P_L'}}^{\dagger}{V}_{B,-1}^{(L)} & 0 \\
%%  \multicolumn{2}{c}{0} \\
%%  \multicolumn{2}{c}{\vdots} \\
  \multicolumn{2}{c}{0} \\
  0 & -{\overleftarrow{P_R'}}{V}_{A,1}^{(R)} \\
  0 & {P_R'}^{\dagger}{V}_{B,0}^{(R)} \\
  \end{pmatrix},  
\end{equation}
\end{widetext}
where $\left\{\Psi_C^{(l)}\right\}$ is the set of wavefunctions for the
central section, for $l=1\dots{}N_{\rm pro}^{(L)}+N_{\rm pro}^{(R)}$, stored
as columns. 
In the above, we used Eqs.\ (\ref{simple-ul}), (\ref{simple-br}). 

The rank of the main matrix in Eq.\ (\ref{kramsys}) is equal to
$N_{\rm pro}^{(L)}+2N_CM+N_{\rm pro}^{(R)}$, where $M$ denotes the number of vertical
sections, each containing (up to) $2N_C$ lattice sites.
Although the shape of the central section is --- in principle --- irregular,
a~site-numbering scheme analogous to that used for the leads can be applied.
In case of a~considerable number of disconnected lattice sites, an additional 
enumeration for the central part can be applied to eliminate the matrix
elements corresponding to disconnected sites and to reduce the rank of the
main matrix. 

For the total number of sites $2N_CM<5\times{}10^5$, such as in the numerical
examples presented in the main text, one can use standard linear-algebra
software to solve the linear system given by Eq.\ (\ref{kramsys}) numerically
and find the unknowns, including all elements of the scattering matrix $S$ 
(\ref{smatdef}).
We used the double precision LAPACK routine {\tt zgbsv}, see
Ref.\ \cite{lapack99}, which employs the band storage scheme for the
main matrix with $2N_C-1$ subdiagonals and $2N_C-1$ superdiagonals.
Assuming that the proportions of the middle section are not very different
from those of the square, i.e., $N_C\sim{}M\sim{}\sqrt{N_{\rm tot}}$
(with the total number of sites $N_{\rm tot}$) we can estimate the
computational complexity to be ${\cal O}(N_{\rm tot}^2)$, which is equivalent to
the complexity of the more commonly used recursive Green’s function method
\cite{Lew13}. 

Since the normal modes in the left and right leads are calculated
independently, we can validate the final numerical output by checking
the unitarity condition for the scattering matrix,
$SS^\dagger{}=S^\dagger{}S=\mathbb{I}$.
The deviation from unitarity is quantified by
\begin{equation}
  \varepsilon_S =
%% \mbox{max}_{1\leqslant{}l,l'\leqslant{}N_{\rm pro}^{(L)}+N_{\rm pro}^{(R)}}
%%  \left|\left(S^{\dagger}S\right)_{ll'}-\delta_{ll'}\right|.
  \mbox{max}\left\{
  \left|\left(S^{\dagger}S\right)_{ll'}-\delta_{ll'}\right|,\ 
  {1\leqslant{}l,l'\leqslant{}N_{\rm pro}^{(L)}\!+\!N_{\rm pro}^{(R)}}
  \right\}. 
\end{equation}
In the numerical problems considered here, the above does not
exceed $\varepsilon_{S}\lesssim{}10^{-6}$; the maximal value is
approached for the half-disk system, for which the number of propagating
modes in the right (i.e., the wider) lead is
$52\leqslant{}N_{\rm pro}^{(R)}\leqslant{}537$
for the energy range of $0.1\leqslant{}(E+V_0)/t_0\leqslant{}0.9$.

%% Patrz plik: "data1/gcor700x240R200r50dv0.50am0.01ef-ALL.dat"

If more than two leads contact the central system, a~generalization
is straightforward, provided we have a~group of (one or more) parallel
left leads and a~group of parallel right leads.
In such a~case, the block structure of Eq.\ (\ref{kramsys}) is preserved, 
and it is only necessary to modify the contents of the blocks connecting
the central section and the leads accordingly. 
For some more complex cases, relevant graph algorithms are implemented
in the KWANT package \cite{Gro14}. 

For much larger systems, the direct lattice approach presented above must
be replaced by the method truncating the wavefunction within 
orthogonal polynomials, such as recently implemented in the KITE software 
\cite{Joa20}.

%%%%%%%%%%%%%%%%%%%%%%%%%%%%%%%%%%%%%%%%%%%%%%%%%%%%%%%%%%%%%%%%%%%%%%%%%%%%%%
%%%%%%%%%%%%%%%%%%%%%%%%%%%%%%%%%%%%%%%%%%%%%%%%%%%%%%%%%%%%%%%%%%%%%%%%%%%%%%

%%%%%%%%%%%%%%%%%%%%%%%%%%%%%%%%%%%%%%%%%%%%%%%%%%%%%%%%%%%%%%%%%%%%%%%%%%%%%%
%%%%%%%%%%%%%%%%%%%%%%%%%%%%%%%%%%%%%%%%%%%%%%%%%%%%%%%%%%%%%%%%%%%%%%%%%%%%%%


\begin{thebibliography}{}
\bibitem{diskfoo}
For the disk geometry, the width $W$ needs to be replaced with the inner
lead circumfence $2\pi{}R_{{\rm i}}$.
\bibitem{planewfoo}
Strictly speaking, these solutions take the form of linear combinations of
two plane waves (with positive and negative $k_y$), producing standing waves
in the $y$ direction, but this detail is irrelevant to the calculations
presented below.
\bibitem{valleyfoo}
It should be noted that Eq.\ (\ref{diraceq}) is used for the $K$-valley
neighborhood in the dispersion relation; for the $K’$-valley the matrix
$\sigma_y$ should be replaced by $\sigma_y^\star=-\sigma_y$.
\bibitem{matchfoo}
Similar to the Schr\"{o}dinger equation case, the wave function matching
conditions are a~consequence of charge conservation.
To obtain the current density operator, in the Hamiltonian operator on the
left side of Eq.\ (\ref{diraceq}),
$H=v_F\,{\boldsymbol p}\cdot{\boldsymbol\sigma}+V(x)$, we substitute
${\boldsymbol p}\rightarrow{\boldsymbol p}+e{\boldsymbol A}$, and then
differentiate the resulting operator $H({\boldsymbol A})$ with respect
to the vector potential ${\boldsymbol A}$. In effect, we obtain
${\boldsymbol j}=ev_F{\boldsymbol\sigma}$, and therefore the continuity of
the current corresponds to the continuity of the components of the wave
function (operator ${\boldsymbol j}$ does not involve differentiation over
coordinates).
\bibitem{nmodfoo}
It can be further approximated as $N_{\rm approx}=2KW_{\rm infty}/\pi$
with $K=|E+V_{\infty}|/\hbar{}v_F$, giving $N_{\rm approx}\simeq{}130$ for $E=0$
and $V_{\infty}=t_0/2$; the actual number of propagating
modes $N_{\rm open}^{(\infty)}\geqslant{}N_{\rm approx}$ (if 
$|E+V_{\infty}|\lesssim{}t_0$) due to the trigonal warping.
\end{thebibliography}
\end{document}